\documentclass[12pt]{article}
\usepackage[nosort]{cite}
\usepackage[usenames, dvipsnames]{xcolor}
\usepackage{graphicx}
\usepackage{multicol}
\usepackage{amsfonts}
\usepackage{amssymb}
\usepackage{amsmath}
\usepackage{afterpage}
\usepackage{setspace}
\usepackage{verbatim}
\usepackage{color}
\usepackage{longtable}
\usepackage{caption}
\usepackage{heck}
\usepackage{float}
\usepackage{slashed}
\usepackage[utf8]{inputenc}
\usepackage{braket}
\usepackage{subcaption}
\usepackage{epsfig}
\usepackage{enumerate}
\usepackage{epstopdf}
\usepackage[enableskew, vcentermath]{youngtab}
\usepackage{adjustbox}
\usepackage{multirow}
\usepackage{tikz}
\usepackage[margin=1in]{geometry}
\usepackage{titletoc}
\usepackage{hyperref}
\usepackage[percent]{overpic}
\usepackage{gensymb}
\usepackage{mathtools}
\usepackage{tikz-cd}%
\usepackage[utf8]{inputenc}
\usepackage{epsf}
\usepackage{colortbl}
\usepackage{amsmath}
\usepackage{amsfonts}
\usepackage{amssymb}
\usepackage{graphicx}
\usepackage{color}
\usepackage{psfrag}
\usepackage{cite}
\usepackage{hyperref}
\usepackage{mathrsfs}
\usepackage{tikz}
\usepackage{tikz-feynman}
\usepackage{slashed}
\usepackage{ifpdf}
\usepackage{mdframed}
\mdfdefinestyle{statementstyle}{
   frametitlerule = true,
   frametitlefont = {\normalfont\bfseries},
   frametitlebackgroundcolor = gray!10,
}
\mdtheorem[style=statementstyle]{statement}{Definition}
\providecommand{\U}[1]{\protect\rule{.1in}{.1in}}
\pdfoutput=1
\newsavebox{\mysavebox}

\hypersetup{colorlinks,citecolor=black,filecolor=black,linkcolor=black,urlcolor=black}
\usetikzlibrary{decorations.markings}

\numberwithin{equation}{section}

\usetikzlibrary{chains}
\allowdisplaybreaks
\tikzset{node distance=2em, ch/.style={circle,draw,on chain,inner sep=2pt},chj/.style={ch,join},every path/.style={shorten >=4pt,shorten <=4pt},line width=1pt,baseline=-1ex}

\newcommand{\ba}{\begin{eqnarray}}
\newcommand{\ea}{\end{eqnarray}}

\newcommand{\be}{\begin{equation}}
\newcommand{\ee}{\end{equation}}

\newcommand{\eq}[1]{(\ref{#1})}

\tikzstyle{startstop} = [rectangle, rounded corners, minimum width=3cm, minimum height=1cm,text centered, draw=black, fill=blue!10]
\tikzstyle{startstop} = [rectangle, rounded corners, minimum width=3cm, minimum height=1cm,text centered, draw=black, fill=blue!10]
\tikzstyle{io} = [trapezium, trapezium left angle=70, trapezium right angle=110, minimum width=3cm, minimum height=1cm, text centered, draw=black, fill=blue!30]
\tikzstyle{process} = [rectangle, minimum width=3cm, minimum height=1cm, text centered, draw=black, fill=orange!30]
\tikzstyle{decision} = [diamond, minimum width=3cm, minimum height=1cm, text centered, draw=black, fill=green!30]
\tikzstyle{arrow} = [thick,->,>=stealth]
\tikzset{->-/.style={decoration={
  markings,
  mark=at position #1 with {\arrow[scale=2.4]{>}}},postaction={decorate}}}
\makeatletter \@addtoreset{equation}{section} \makeatother

\renewcommand{\[}{\left[}

\colorlet{darkblue}{blue!70!black}
\colorlet{darkgreen}{green!70!black}

\hypersetup{
colorlinks=true,
linkcolor=blue,
citecolor=magenta,
}

\begin{document}

\date{December 2022}

\title{IIB Explored: Reflection 7-Branes}

\institution{LMU}{\centerline{$^{1}$Arnold Sommerfeld Center for Theoretical Physics, LMU, Munich, 80333, Germany}}
\institution{PENN}{\centerline{$^{2}$Department of Physics and Astronomy, University of Pennsylvania, Philadelphia, PA 19104, USA}}
\institution{PENNmath}{\centerline{$^{3}$Department of Mathematics, University of Pennsylvania, Philadelphia, PA 19104, USA}}
\institution{HARVARD}{\centerline{$^{4}$Department of Physics, Harvard University, Cambridge, MA 02138, USA}}

\authors{
Markus Dierigl\worksat{\LMU}\footnote{e-mail: \texttt{m.dierigl@lmu.de}},
Jonathan J. Heckman\worksat{\PENN,\PENNmath}\footnote{e-mail: \texttt{jheckman@sas.upenn.edu}},\\[4mm]
Miguel Montero\worksat{\HARVARD}\footnote{e-mail: \texttt{mmontero@g.harvard.edu}}, and
Ethan Torres\worksat{\PENN}\footnote{e-mail: \texttt{emtorres@sas.upenn.edu}}
}

\abstract{\noindent The Swampland Cobordism Conjecture successfully predicts the supersymmetric spectrum of 7-branes
of IIB / F-theory. Including reflections on the F-theory torus, it also predicts the existence
of new non-supersymmetric objects, which we dub reflection 7-branes (R7-branes).
We present evidence that these R7-branes only exist at strong coupling. R7-branes serve as end of the world branes for 9D theories obtained from type IIB asymmetric orbifold and Dabholkar-Park orientifold backgrounds, and an anomaly inflow analysis suggests the existence of a gapless Weyl fermion, which would have the quantum numbers of a goldstino. Using general arguments, we conclude that different kinds of branes are able to end on the R7, and accounting for their charge requires exotic localized degrees of freedom, for which the simplest possibility is a massless 3-form field on the R7-brane worldvolume. We also show how to generalize the standard F-theory formalism to account for reflections.}

{\small \texttt{\begin{flushright}   LMU-ASC 31/22 \end{flushright} }}

\maketitle

\setcounter{tocdepth}{2}

\tableofcontents

\newpage

\section{Introduction}

Dualities provide an important handle on non-perturbative phenomena in quantum
gravity. This has made it possible to extract a number of robust features of
strongly coupled systems which would otherwise be computationally intractable.

That being said, the strongest evidence for these dualities comes from
supersymmetric systems where it is possible to perform precision tests.
One of the major stumbling blocks in recent years has
centered on how to study non-perturbative phenomena without the crutch of
supersymmetry. Indeed, there is (as yet) no evidence for supersymmetry in our
Universe, and so it is natural to seek out complementary methods to address
non-perturbative phenomena in string theory / quantum gravity.

A quite promising recent development in this direction is the advance of new
topological methods to address such questions. This includes, for example, an
improved understanding on the topological structure of generalized symmetries
in quantum field theory \cite{Gaiotto:2014kfa}. In gravity, of course, one does not expect global symmetries, in part because topology can also fluctuate. The Cobordism Conjecture \cite{McNamara:2019rup} encapsulates this general feature through the requirement that the bordism group of quantum gravity is actually trivial:
\begin{equation}
\Omega_{k}^{\text{QG}}=0.
\end{equation}
The power of the Cobordism Conjecture is that at low energies, one often considers bordisms which retain specific symmetry structures $\mathcal{G}$ for which $\Omega^\mathcal{G}_k$ is non-zero. A non-vanishing value of $\Omega_{k}^{\mathcal{G}}$ thus \textit{predicts} that there must be a corresponding object in the spectrum of quantum gravity which trivializes this generator in $\Omega_{k}^{\text{QG}}$.

In many cases these correspond to known supersymmetric defects; but it is especially interesting when the Cobordism Conjecture predicts new, non-supersymmetric objects. For instance, in \cite{Montero:2020icj}, supersymmetric objects predicted by the Cobordism Conjecture were used to make a Swampland prediction on the rank of 8D and 9D supersymmetric theories, which matched string theory calculations.  See \cite{Montero:2020icj, Dierigl:2020lai, Buratti:2021fiv, Debray:2021vob, Blumenhagen:2021nmi,
Angius:2022aeq, Blumenhagen:2022mqw, Angius:2022mgh, Blumenhagen:2022bvh} for further applications of the Cobordism Conjecture.

The case we will study in this paper corresponds to a new kind of non-supersymmetric
7-brane in type IIB string theory, different from the $[p,q]$ 7-branes familiar from IIB string theory.
This brane is the cobordism defect associated to a duality bundle not contained in $SL(2,\mathbb{Z})$.

While it is common to assert that the duality group of type IIB string theory
is simply $SL(2,\mathbb{Z})$, accounting for reflections of the F-theory torus and taking into account their action on fermions leads to the full duality group being the $\mathsf{Pin}^+$ double cover of $GL(2,\mathbb{Z})$ which we denote as $GL^{+}(2,\mathbb{Z})$ (see \cite{Tachikawa:2018njr, Debray:2021vob}).\footnote{The
elements of $GL(2,\mathbb{Z})$ consist of
$2\times2$ matrices with integer entries and determinant $\pm1$. All other
values of the determinant are excluded because the inverse (in
$GL(2,\mathbb{R})$) would not have integer entries.} Taking into account the correlation between spacetime
spin and duality transformations, in the 10D spacetime, a non-trivial duality bundle
will have a structure group of the form $\left(\text{Spin}\times GL^{+}(2,\mathbb{Z})\right)  /%
\mathbb{Z}
_{2}$.

Reference \cite{BIGKAHUNA} computes the bordism groups for these
different choices of duality bundle. For the case of $SL(2,\mathbb{Z})$ and $Mp(2,\mathbb{Z})$
duality bundles, we find that there are corresponding supersymmetric F-theory
backgrounds which serve as generators for the corresponding bordism groups.
This is a remarkably precise test of the Cobordism Conjecture, and also
provides strong evidence that little is missing in our understanding of
supersymmetric F-theory backgrounds.

But in the case of the full duality group $GL^{+}(2,\mathbb{Z})$, where we do
not insist on supersymmetry, we find a surprise. The computations of
\cite{BIGKAHUNA} show that
\begin{equation}\Omega_{1}%
^{\text{Spin-}GL^+(2,\mathbb{Z})}=\mathbb{Z}_2\times \mathbb{Z}_2,\end{equation} where only one of the two $\mathbb{Z}_2$ factors is killed by a known supersymmetric 7-brane (an  F-theory singularity with $\tau = i$ such as a type $III$ or type $III^\ast$ fiber). To kill the other one, we need to introduce a new species of 7-brane\footnote{See also the brief remark on p.9 of \cite{Distler:2009ri}.}. It is non-supersymmetric and, because it kills the class of F-theory specified by monodromy consisting of a reflection along the a- or b-cycle of the F-theory torus, we call it the \textquotedblleft reflection
7-brane\textquotedblright\ (R7-brane). The R7-branes can be regarded as providing a boundary condition for the type IIB Asymmetric Orbifold / Dabholkar Park backgrounds constructed in \cite{Dabholkar:1996pc, Hellerman:2005ja, Aharony:2007du}.  An additional
outcome of \cite{BIGKAHUNA} is that all of the odd bordism groups $\Omega
_{k}^{\text{Duality}}$ can be accounted for by backgrounds with
such R7-branes (as well as more standard supersymmetric objects)
included \cite{BIGKAHUNA}.

Our goal in this paper will be to establish some basic properties of such
R7-branes. Far from these branes, we can work at weak string coupling,
but general arguments imply that we expect the dilaton to approach an order one value near the
core of these branes. Based on their monodromy action, these branes come in
two general types which we refer to as the $\Omega$-brane and the
$\mathsf{F}_{L}$-brane, which are related by S-duality. We find that these
branes do not preserve supersymmetry, but can still form supersymmetric bound
states. For example, the combined monodromy from $\Omega$ and $\mathsf{F}_{L}$
is the same as that of an $I_{0}^{\ast}$ Kodaira fiber with gauge algebra
$\mathfrak{so}(8)$ as obtained from a configuration of 4 D7-branes and an
O7$^{-}$-plane. The absence of supersymmetry means that these branes may be unstable, and eventually \textquotedblleft
explode\textquotedblright. Whether they do this or not, we can still use constraints
from topology to study many of their properties.

To gain more insight into the microscopic structure of these branes, we also
turn to their anomalies. Such R7-branes arise as boundaries of supersymmetric asymmetric orbifold
backgrounds of the type considered in \cite{Hellerman:2005ja, Aharony:2007du}
known as AOB\ backgrounds. We can use anomaly inflow arguments from
the bulk to these 8D objects to further constrain their worldvolume degrees of
freedom. We find that anomalies can be cancelled in this way, most naturally if the worldvolume degrees of freedom contains a complex 8D Weyl fermion (although there are other possibilities compatible with anomaly cancelation). If the anomaly is indeed cancelled by an 8D Weyl fermion, its natural interpretation would be a goldstino arising from the spontaneous supersymmetry breaking on the R7-brane.

We also study brane probes of the worldvolume
theory living on these R7-branes. These probe branes can encircle the R7-brane, and owing to the presence of non-trivial branch cuts / monodromy, can form ``lasso'' configurations in which a junction of objects encircles the R7-brane and the excess charge extends to infinity.\footnote{In this sense, these R7-branes are a generalization of \textquotedblleft Alice strings\textquotedblright\ in 4D gauge field theory \cite{Schwarz:1982ec, Schwarz:1982zt}.} Shrinking this lasso to zero size then implies the existence of objects which can end on the R7-brane. As is familiar from the case of D7-branes, the R7-branes must have the appropriate worldvolume fields to allow differently charged objects to end. Due to the different symmetry properties, we learn that we must have what looks like a massless 3-form field in the 8D worldvolume of the R7-brane, in addition to ordinary gauge fields. This is rather different from what appears in the worldvolume theory of a standard D-brane, for example, and underscores the novel structure of the R7-brane.

Finally,  we also show how these objects can be combined to yield
consistent F-theory backgrounds. In F-theory it is natural to geometrize the
$SL(2,\mathbb{Z})$ duality bundle (and its $Mp(2,\mathbb{Z})$ cover) in terms
of a family of spacetime varying elliptic curves (as well as generalizations to genus one curves).
In the broader context of $GL(2,\mathbb{Z})$ bundles (and its $GL^{+}(2,\mathbb{Z})$ cover), we must
instead allow for possible orientation reversals. This in turn requires us to
consider a doubled elliptic fibration in which we consider a pair of two elliptic
fibrations with possible orientation reversals. We mainly illustrate this by
way of example, but it is clear that this provides a general template for
capturing the topological structure of F-theory backgrounds with R7-branes included.

The rest of this paper is organized as follows. We begin in section \ref{sec:AFAR} by
briefly reviewing the arguments from \cite{BIGKAHUNA} for the existence of
R7-branes in IIB\ / F-theory. Using the asymptotic monodromic action specified
by these branes, we argue that they do not preserve supersymmetry, but can
nevertheless form S-dual pairs with supersymmetry preserving monodromies. Section \ref{sec:PROBE} discusses brane probes of R7-branes and establishes that various objects can end on R7-branes.
In Section  \ref{sec:ANOMO} we use symmetric mass generation arguments to conclude that the R7-brane requires a strong coupling description, and use anomaly inflow and RR-charge conservation to argue that
the worldvolume theory of these branes to be likely given massless fermions and $k$-form potentials. In section \ref{sec:DOUBLE} we formalize some of these considerations by introducing a
prescription for doubled elliptic fibrations which can accommodate reflections
on the F-theory torus. We present our conclusions and future direction in section \ref{sec:CONC}.
Appendix \ref{app:weierstrass} contains further details on elliptic curves and Weierstrass $\wp$-functions.
Some additional details on $\mathsf{Pin}^{\pm}$ structures of Klein bottles are discussed in Appendix \ref{app:kb}, while Appendix
\ref{app:mlift} shows how (worldsheet) chiral fermion parities in string theory arise from M-theory reflections. Appendix \ref{app:CS1loop}
presents a more general analysis of anomalies of various IIB backgrounds under circle reduction. The main results
presented here were announced in the talks \cite{MonteroTalk, HeckmanTalk}.

\section{R7-Branes from Afar} \label{sec:AFAR}

In this section we explain how the Cobordism Conjecture leads to the
prediction of R7-branes. With this in place, our task in the remaining
sections will be to determine some basic properties of these objects. We begin
by briefly reviewing the Cobordism Conjecture, as well as its
implications for type IIB\ duality, and the prediction that R7-branes exist.
With this in place, we show that supersymmetry is broken by these branes. We
then determine some basic features of such branes, including possible annihilation channels.

\subsection{Bordisms and Dualities}

\begin{figure}[ptb]
\centering
\includegraphics[width = 0.4 \textwidth]{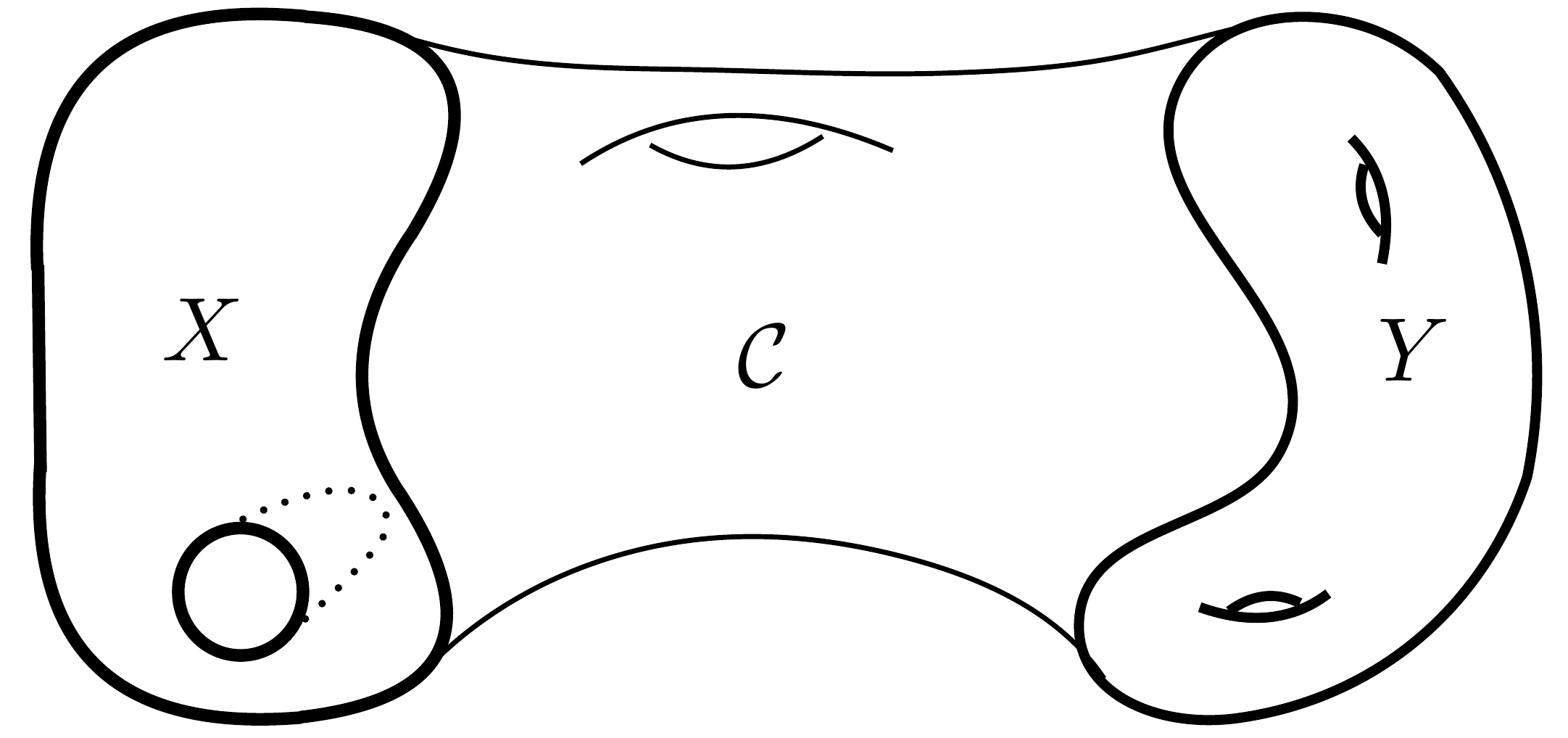}
\caption{Depiction of a bordism $\mathcal{C}$ between two manifold $X$ and $Y$ of one dimension less. The physical interpretation of the cobordism is a topology-changing transition described by a smooth spacetime, that can be treated in the low-energy supergravity. If the underlying theory includes additional structures (orientation, spin, a principal bundle\ldots), we demand these can be extended from $X$ and $Y$ to $\mathcal{C}$.}
\label{f1}
\end{figure}

As explained in the Introduction, the Cobordism Conjecture represents an extension of the conjecture that there are no global symmetries in quantum gravity to situations involving dynamical topology change, illustrated in Figure \ref{f1}. In physical terms, bordisms describe interpolating spaces that can be viewed as off-shell
configurations of the gravitational path integral. The existence of these configurations means that the space $X$ in Figure \ref{f1} is not ``conserved'', and it may fluctuate to some other space $Y$ via a process that, although off-shell, can be completely described by the low-energy supergravity theory, since it does not involve singular spaces or infinite energy densities. However, the bordism class $[X]=[Y]$ may be conserved, at least by the topology-changing processes that can be described by the low-energy supergravity. In practice, for most supergravity theories of interest,
there are many non-vanishing bordism groups, and so as a consequence, the
Cobordism Conjecture predicts that there must be corresponding
defects which trivialize these classes in quantum gravity\footnote{Strictly speaking, we can either break or gauge the symmetry in question. In the latter case, this amounts to saying that the background we started with does not make sense in quantum gravity to begin with. This can happen if e.g., there is a probe leading to an anomaly in this background. A concrete example is the class of $\mathbb{RP}^4$ in M-theory with vanishing $G_4$ flux, which is inconsistent due to an anomaly in the M2-brane wordlvolume \cite{Witten:2016cio}. All the cases considered in this paper correspond to IIB backgrounds which we believe make sense, so the only option allowed by the Cobordism Conjecture is that we have a
suitable defect killing the bordism class in question.}. These objects may be very singular from the low-energy point of view, and their inclusion would be completely unjustified from a low-energy EFT analysis; we only know they must make sense due to Swampland principles.

With this in mind, the basic idea to apply the Cobordism Conjecture
is to first identify the bordism groups relevant to the theory under consideration, and then use it to predict new defects.
Our interest in this work will be the specific case of dualities for type
IIB string theory. Of course, to carry out this analysis, we must first
provide a precise specification of the class of manifolds in which IIB string theory, or more precisely, its low-energy approximation IIB supergravity, makes sense. A first approximation would be to focus on Spin manifolds, since the theory contains fermions. Therefore, both $X$ and $Y$ in Figure \ref{f1}, as well as the bordism $\mathcal{C}$, must carry a Spin structure.
For the purposes of this paper however, we wish to also consider duality bundles for the type IIB supergravity theory.
At the level of the IIB\ supergravity action, the duality symmetry is sometimes presented as
$SL(2,\mathbb{R})$, which is further broken to $SL(2,\mathbb{Z})$ once one
takes into account flux quantization and branes. A famous fact of F-theory
\cite{Vafa:1996xn, Morrison:1996na, Morrison:1996pp} is that this duality symmetry can be interpreted geometrically as the
group of large diffeomorphisms of a torus, its complex structure $\tau$
corresponding to the type IIB\ axio-dilaton. Including the action on fermionic
degrees of freedom such as the dilatinos and gravitinos, this is refined to
the metaplectic cover $Mp(2,\mathbb{Z})$ \cite{Pantev:2016nze}.\footnote{This is
analogous to the group $Spin(n)$ being a double cover of the group $SO(n)$.}
We note that including this Spin structure is implicitly handled in most
F-theory constructions since they typically always deal with Calabi-Yau spaces.

But even $Mp(2,\mathbb{Z})$ is not the end of the story. Owing to M/
F-theory duality, or directly from a study of IIB perturbative symmetries \cite{Tachikawa:2018njr},
reflections of the F-theory torus also constitute symmetries of the theory.
In purely type IIB terms, such reflections correspond to acting by worldsheet
parity $\Omega$ or leftmoving spacetime fermion parity $\mathsf{F}_L \equiv (-1)^{F_{L}}$, depending on whether the reflection takes place on the a-cycle or b-cycle. At the level of bosonic degrees of freedom, this means that we must enlarge the
duality group to $GL(2,\mathbb{Z})$, namely we also allow determinant $-1$
matrices. Including fermionic degrees of freedom, we are finally led to the
$\mathsf{Pin}^{+}$ cover of $GL(2,\mathbb{Z})$ denoted by $GL^{+}(2,\mathbb{Z})$. Since the $\mathbb{Z}_2$ subgroup of this is identified with spacetime fermion number, the structure we will demand in our IIB backgrounds is that the fermions live in sections of a
\begin{equation}
\frac{\text{Spin}\times GL^{+}(2,\mathbb{Z})}{\mathbb{Z}_2}
\end{equation}
bundle over spacetime, which we dub a Spin-$GL^{+}(2,\mathbb{Z})$ structure \cite{Debray:2021vob, BIGKAHUNA}.
We will therefore demand that our IIB backgrounds in Figure \ref{f1} have a Spin-$GL^{+}(2,\mathbb{Z})$ structure in both $X$, $Y$ and $\mathcal{C}$, and that the Spin-$GL^{+}(2,\mathbb{Z})$ structure in $\mathcal{C}$ restricts smoothly to $X$ and $Y$.  Excluding reflections, we can of course restrict to the special case of
Spin-$Mp(2,\mathbb{Z})$ bundles, and this is the case which has been studied
in much of the F-theory literature.

Reference \cite{BIGKAHUNA} computes the bordism groups for both
Spin-$Mp(2,\mathbb{Z})$ and Spin-$GL^{+}(2,\mathbb{Z})$ structure manifolds. Quite
remarkably, it turns out that for all odd $k$, the generators in the Spin-$Mp(2,\mathbb{Z})$ case correspond to known
objects of supersymmetric F-theory backgrounds.\footnote{The case of $k$ even
is essentially captured by purely gravitational defects of the type already found in
\cite{McNamara:2019rup}.} To give some examples, in the case $k=1$, we obtain the expected
7-branes of F-theory; for $k=3$, this correctly captures 7-branes wrapped on
rigid curves at fixed axio-dilaton, namely the non-Higgsable clusters of
reference \cite{Morrison:2012np}; and for $k=5$ we correctly predict orientifold
3-planes and their strongly coupled generalizations known as S-folds (see e.g, \cite{Garcia-Etxebarria:2015wns, Aharony:2016kai}).

In the case of Spin-$GL^{+}(2,\mathbb{Z})$ bordisms, a similarly striking match
is obtained, with one notable exception. It turns out that there is one new
bordism generator of $\Omega_{1}^{\text{Spin-}GL^+(2,\mathbb{Z})}$ which is not generated by a
known supersymmetric background. That being said, combining this single new
object with all of the (supersymmetric) generators of Spin-$Mp(2,\mathbb{Z})$
bordism is enough to completely fill out the full list of objects predicted by
the Cobordism Conjecture.

This begs the question: what is the new generator of $\Omega_{1}^{\text{Spin-}GL^{+}(2,\mathbb{Z})}$? To answer this, it is helpful to
compare with the supersymmetric Spin-$Mp(2,\mathbb{Z})$ case. As explained in \cite{McNamara:2021cuo},
for any structure group $\mathcal{G}$, the bordism group
$\Omega^{\text{Spin-}\mathcal{G}}_1$ is given by $\mathrm{Ab}[\mathcal{G}]$, the Abelianization of $\mathcal{G}$.
Alternatively, one can use the results in \cite{BIGKAHUNA} (see also \cite{Dierigl:2020lai}) to obtain:
\begin{align}
\Omega_{1}^{\text{Spin-}Mp(2,\mathbb{Z})}  &  =%
\mathbb{Z}
_{3}\times%
\mathbb{Z}
_{8}\\
\Omega_{1}^{\text{Spin-}GL^{+}(2,\mathbb{Z})}  &  =%
\mathbb{Z}
_{2}\times%
\mathbb{Z}
_{2}.
\label{mhht}\end{align}
In the case of the $\Omega_{1}^{\text{Spin-}Mp(2,\mathbb{Z})}$ bordism
generators, the  F-theory background is realized as the boundary
of the supersymmetry preserving orbifolds
\begin{equation}\left(  \mathbb{C}\times
T^{2}\right)  /%
\mathbb{Z}
_{3}\quad \text{and}\quad \left(  \mathbb{C}\times T^{2}\right)  /%
\mathbb{Z}
_{4}.\end{equation} These backgrounds describe a 7-brane with constant axio-dilaton (see e.g., \cite{Dasgupta:1996ij}).
In the $%
\mathbb{Z}
_{3}$ case we have $\tau=\exp(2\pi i/3)$ and Kodaira fiber of type IV$^{\ast}$ (namely
an $\mathfrak{e}_6$ gauge algebra). In the $%
\mathbb{Z}
_{8}$ case we have $\tau=\exp(2\pi i/4)$ and Kodaira fiber of type III$^{\ast
}$ (namely an $\mathfrak{e}_{7}$ gauge algebra). So these 7-branes are the supersymmetric
defects killing the corresponding bordism clases, as advertised.

Turning next to the case of the $\Omega_{1}^{\text{Spin-}GL^{+}(2,\mathbb{Z}%
)}$ bordism generators, one of the $%
\mathbb{Z}
_{2}$ factors is simply a remnant of the $\left(  \mathbb{C}\times
T^{2}\right)  /%
\mathbb{Z}
_{4}$ background. Compared with its $Mp(2,\mathbb{Z})$ counterpart, it is a
smaller group now because due to reflections, more identifications between
backgrounds are available (more group elements are commutators). On the other hand, the $%
\mathbb{Z}
_{3}$ group is completely absent and is instead replaced by another $%
\mathbb{Z}
_{2}$ generator.\footnote{As explained in \cite{BIGKAHUNA}, the $%
\mathbb{Z}
_{3}$ is trivial once reflections are included because we can annihilate odd
order elements of $\Omega_{1}^{\text{Spin-}Mp(2,\mathbb{Z})}$. Again, this can also be understood as the fact that the spin lift of the $U$ generator can be written as a commutator in $\text{Spin}$-$GL^{+}(2,\mathbb{Z})$,
but not in  $\text{Spin}$-$Mp(2,\mathbb{Z})$.} In terms of
the local coordinates $(\theta;a,b)$ of an $S^{1}$ and a torus, this monodromy
acts via a reflection on the $a$-cycle of the torus. Working in a coordinate
system where $\theta$ is $2\pi$-periodic, the transformation rule
is:
\begin{equation}
(\theta;a,b)\mapsto(\theta+2\pi;-a,b).
\end{equation}
The other S-dual reflection on the $b$-cycle does not generate a distinct
bordism class since we can of course permute the roles of the $a$- and $b$-cycles.

From these basic considerations, we conclude that the Cobordism
Conjecture is making a firm prediction:\ there is a new kind of real codimension
two object in type IIB\ backgrounds as associated with a monodromic action by
a reflection. We will dub these objects \emph{reflection 7-branes}, or R7-branes for short. An important difference is that, unlike for the  branes described above, there is no way to construct these objects as an orbifold of a smooth IIB or F-theory configuration configuration; we will explain why in more detail when we study the supersymmetry of the background in section \ref{ssec:SUSYBREAK}.

 Let us now determine some basic properties of the reflection 7-branes. We denote by $\mathsf{F}_{L}$ the 7-branes associated with an
$a$-cycle reflection and $\Omega$ the 7-brane associated with a $b$-cycle
reflection. In terms of their representatives in $GL(2,\mathbb{Z})$ (i.e.,
neglecting their action on spinors) these are captured by the monodromy
matrices:%
\begin{equation}
M_{\mathsf{F}_{L}}=\left[
\begin{array}
[c]{cc}%
-1 & \\
& +1
\end{array}
\right]  \text{, \ \ }M_{\Omega}=\left[
\begin{array}
[c]{cc}%
+1 & \\
& -1
\end{array}
\right]  .
\end{equation}
Note in particular that these are not elements of $SL(2,\mathbb{Z})$. As reviewed
in \cite{Tachikawa:2018njr}, the action on the bosonic IIB\ fields is:
\begin{equation}%
\begin{tabular}
[c]{|l|l|l|l|l|l|l|}\hline
& $g$ & $\phi$ & $C_{0}$ & $C_{4}$ & $C_{2}$ & $B_{2}$\\\hline
$\Omega$ & $+1$ & $+1$ & $-1$ & $-1$ & $+1$ & $-1$\\\hline
$\mathsf{F}_{L}$ & $+1$ & $+1$ & $-1$ & $-1$ & $-1$ & $+1$\\\hline
\end{tabular}
\text{.}%
\end{equation}
In particular the action on the axio-dilaton $\tau=C_{0}+ie^{-\phi}$
is, for both monodromies:%
\begin{equation}
\tau\rightarrow-\overline{\tau}\text{,}%
\end{equation}
which is not holomorphic.

Note, however, that the combined product $M_{\mathsf{F}_{L}}M_{\Omega}$ is just
minus the identity. In fact, this is precisely the monodromy obtained from
encircling a Kodaira fiber of type $I_{0}^{\ast}$ which is locally modelled as
a $D_{4}$-type singularity (i.e., the quotient of $\mathbb{C}^2$ by the
binary dihedral group with eight elements):%
\begin{equation}
M_{\mathsf{F}_{L}}M_{\Omega}=M_{I_{0}^{\ast}}=\left[
\begin{array}
[c]{cc}%
-1 & \\
& -1
\end{array}
\right]  \text{.}%
\end{equation}
In perturbative type IIB\ string theory, this is realized by 4 D7-branes and
an O7$^{-}$-plane filling $\mathbb{R}^{7,1}$ and sitting at the same point in
the transverse directions. As is well-known, this generates an $\mathfrak{so}%
(8)$ gauge theory with 8D $\mathcal{N}=1$ supersymmetry (16 real
supercharges). So we have shown that the fusion of two R7-branes can yield one of the standard $[p,q]$ 7-brane stacks previously known in the F-theory literature. To put it another way, R7-branes provide a fractionalization of a $D_4$ singularity. As this fractionalization does not correspond to any of the known supersymmetric deformations of the $D_4$ singularity, this is a first hint that the R7-branes are non-supersymmetric.

So far, we have only discussed the  monodromy on bosonic
states, namely we have worked at the level of
$GL(2,\mathbb{Z})$ transformations. But as explained above, the duality group also acts on the
fermionic degrees of freedom, including the dilatinos and gravitinos of
IIB\ supergravity.  For constant axio-dilaton backgrounds, one
can use the transformation rules of \cite{Pantev:2016nze} (see also \cite{Debray:2021vob})
to extract the corresponding action on the fermionic states.
  These reflections can be determined by working either in the M-theory dual picture, or an auxiliary $10+2$ signature spacetime (see also \cite{Heckman:2017uxe, Heckman:2018mxl, Heckman:2022peq}).
 In either formulation, this involves acting on a 2D spinor (supported on the F-theory torus)\ by a
reflection in the a-cycle or b-cycle direction of the F-theory torus, encoded by an action of the form:%
\begin{equation}
R_{i}:\zeta_{2D}\rightarrow\gamma^{i}\,\zeta_{2D},
\end{equation}
where $\zeta_{2D}$ is a 2D\ Majorana spinor and $i=1,2$ indexes the 2D\ Dirac
gamma matrices which satisfy the Euclidean signature Clifford algebra
$\left\{  \gamma^{i},\gamma^{j}\right\}  =2\delta^{ij}$. The overall phase in
this reflection is fixed by the requirement that M-theory has a $\mathsf{Pin}^{+}$
structure, see Appendix \ref{app:mlift} for more details. Fixing an explicit basis compatible with our $\Omega$-brane and
$\mathsf{F}_{L}$-brane actions, we set $\gamma^{1}=\sigma^{1}$ and $\gamma
^{2}=\sigma^{3}$, the standard Pauli matrices. Explicitly, the reflection
action for $\Omega$ and $\mathsf{F}_{L}$ on these 2D\ spinors is \cite{Tachikawa:2018njr}:
\begin{align}
R_{\Omega}  &  :\left[
\begin{array}
[c]{c}%
\zeta^{(1)}\\
\zeta^{(2)}%
\end{array}
\right]  \rightarrow\left[
\begin{array}
[c]{cc}%
0 & 1\\
1 & 0
\end{array}
\right]  \left[
\begin{array}
[c]{c}%
\zeta^{(1)}\\
\zeta^{(2)}%
\end{array}
\right]  =\left[
\begin{array}
[c]{c}%
\zeta^{(2)}\\
\zeta^{(1)}%
\end{array}
\right] \\
R_{\mathsf{F}_{L}}  &  :\left[
\begin{array}
[c]{c}%
\zeta^{(1)}\\
\zeta^{(2)}%
\end{array}
\right]  \rightarrow\left[
\begin{array}
[c]{cc}%
1 & 0\\
0 & -1
\end{array}
\right]  \left[
\begin{array}
[c]{c}%
\zeta^{(1)}\\
\zeta^{(2)}%
\end{array}
\right]  =\left[
\begin{array}
[c]{c}%
\zeta^{(1)}\\
-\zeta^{(2)}%
\end{array}
\right]  .
\end{align}
It is convenient to combine this into a single complex spinor $\zeta
=\zeta^{(1)}+i\zeta^{(2)}$, in terms of which we have:%
\begin{equation}
R_{\Omega}:\zeta\rightarrow i\overline{\zeta}\text{ \ \ and \ \ }%
R_{\mathsf{F}_{L}}:\zeta\rightarrow\overline{\zeta}. \label{reflecto}%
\end{equation}
Comparing with the transformation rules of \cite{Pantev:2016nze} (which restricts attention to $Mp(2,\mathbb{Z})$),
note that there is an additional charge conjugation operation acting on the spinors.

We can also deduce that the R7-branes have non-trivial braiding statistics.
Indeed, observe that the operators $\Omega$ and $\mathsf{F}_{L}$ do not actually commute \cite{Dabholkar:1997zd}.
Instead, we have:
\begin{equation}
\Omega \mathsf{F}_{L}= \mathsf{F}_{R}\Omega=\mathsf{F} \mathsf{F}_{L}\Omega, \label{Braiding}%
\end{equation}
namely there is an insertion of spacetime fermion parity as we attempt to
cross the branes around each other.\footnote{What is the interpretation of this further $\mathsf{F}$-parity defect?
In F-theory terms this corresponds to the insertion of a $dP_{9}$ defect, i.e., a
\textquotedblleft$\frac{1}{2}$K3 surface\textquotedblright. We recall that
these are rational elliptic surfaces given by an elliptic fibration over a
$\mathbb{P}^{1}$. The Weierstrass model for this geometry is of the general
form: $y^{2}=x^{3}+f_{4}x+g_{6}$,
where $f_{4}$ and $g_{6}$ are respectively degree 4 and 6 homogeneous
polynomials in the coordinates of the base $\mathbb{P}^{1}$.}

\subsection{Supersymmetry Breaking} \label{ssec:SUSYBREAK}

We now argue that the R7-branes completely break supersymmetry.
Recall that in a supersymmetric background, there is a non-trivial solution to the (covariant)
Killing spinor equation, specified by the variation of the IIB gravitinos:
\begin{equation}
\delta \Psi_{\mu} = \nabla_{\mu} Q = 0,
\end{equation}
in the obvious notation. Here, $\nabla_{\mu}$ depends on the background fields, including the connection of the duality bundle. The main
claim is that in the presence of an R7-brane, there are no zero mode solutions, and thus supersymmetry is completely broken.

One can already see a potential issue because any candidate 7-brane breaks translations in the directions transverse to its worldvolume, and as
such can retain at most half of the original supersymmetry of the original 32 supercharges of type IIB. Indeed, for the IIB pair of 10D Majorana-Weyl spinors $Q^{(1)}$ and $Q^{(2)}$, only the combination $Q^{(1)} + \Gamma^{9} \Gamma^{8} Q^{(2)}$ survives in the presence of the brane. Precisely because the reflection 7-brane imposes a further transformation on the $Q^{(i)}$'s, the Killing spinor equations become over-constrained, and no zero mode solution can survive.

It is also instructive to see this directly at the level of the monodromy action on any putative zero mode solution.
To do so, consider the monodromy action on 10D supercharges of our system, grouped as a complex supercharge $Q=Q^{(1)}+iQ^{(2)}$.
The R7-branes are codimension-two defects, so we split spacetime as $ \mathbb{R}^{7,1}\times \mathbb{R}^2$.\footnote{Topologically there can be a conical deficit in the presence of our codimension two object.}
The R7-brane sits at $z=0$, and so at the very least, the ten-dimensional Lorentz algebra is broken to
\begin{equation}
\mathfrak{so}(9,1)\,\rightarrow \mathfrak{so}(7,1)\times \mathfrak{so}(2),\label{lor}
\end{equation}
where $\mathfrak{so}(7,1)$ describes Lorentz transformations in $\mathbb{R}^{7,1}$ and $\mathfrak{so}(2)$ describes transformations of the $\mathbb{R}^2$ factor.\footnote{A subtlety here is that because the configuration is non-supersymmetric, it could happen that $\mathfrak{so}(7,1)$ and $\mathfrak{so}(2)$ could actually be spontaneously broken. However, both the geometric construction as well as every 7-brane we know of (including non-BPS 7-branes that we can construct explicitly in string theory \cite{Frau:1999qs, Sen:1999mg}) suggests that these symmetries remain unbroken, and so we will assume that this is the case. This is also compatible with the interpretation of R7-branes as ``end of the world branes'' in AOB and Dabholkar-Park backgrounds (see section \ref{sec:ANOMO}).}

The supercharges $Q^{(i)}$ descend from ten-dimensional spinors, and so decompose into irreducible representations as:
\begin{align}
\mathfrak{so}(9,1) & \supset \mathfrak{so}(7,1) \times \mathfrak{so}(2) \\
\mathbf{16} & \rightarrow \mathbf{8}^+_{1/2}\oplus \mathbf{8}^-_{-1/2}.
\end{align}
Organizing the original spinors into the complexified combination:
\begin{equation}
Q = Q^{(1)} \pm i Q^{(2)},
\end{equation}
under the $\mathfrak{so}(2)$ factor of line (\ref{lor}),
the supercharge $Q$ transforms as:
\begin{equation}
Q \mapsto \exp(i \theta / 2) Q.\label{e23}
\end{equation}
Up to an overall phase (which depends on whether we use the $\Omega$ or $\mathsf{F}_L$ R7-brane), any putative zero mode
undergoes a monodromy as we rotate by $2 \pi$:
\begin{equation}
Q \mapsto \xi \overline{Q},
\end{equation}
for a complex phase $\xi$. One can check that there are no nonvanishing zero modes for equation \eqref{e23} with this boundary condition, so the R7-brane breaks supersymmetry. Notice that this is not the case for monodromies in $Mp(2,\mathbb{Z})$ since in this case the connection inherited from the duality bundle involves a contribution from the spin-connection twisted by the (discrete) connection from the duality bundle. Precisely because $Mp(2,\mathbb{Z})$ transformations have determinant $+1$, any such transformation can be ``undone'' by a suitable transformation in the spin connection, and so there are zero mode solutions to the Killing spinor equations. Said differently, we recover the statement that standard 7-branes are supersymmetric. This is not so for reflections since any putative transformation in $\mathfrak{so}(2)$ would still have determinant
given by $+1$.

We note in passing that \eqref{e23} is a choice, physically motivated but by no means necessary. We could have taken the supercharges to transform in a trivial representation of $SO(2)$; for instance, this is the representation the supercharges transform in when we consider IIB on $S^1$ and interpret $SO(2)$ as the group of isometries of the $S^1$. The choice in equation (\eq{e23}) is equivalent to demanding that the asymptotic $S^1$ is contractible; we take it to be part of the definition of a defect for a bordism class. We will give an independent argument when discussing the R7-brane as a boundary for the AOB background in section \ref{sec:ANOMO}.

\subsection{Brane Annihilation / Recombination}

While it is difficult to directly probe the microscopic degrees of freedom for
these branes, the absence of any preserved supersymmetries means that we should be wary of potential instabilities. One possibility is that the R7-brane ``explodes'', expanding into a growing region without a geometric description, similar to the case of non-supersymmetric orbifold singularities \cite{Adams:2001sv}, where a localized singularity also resolves by expanding out to
the boundaries of the non-compact spacetime. Furthermore, it is generally expected that low-codimension objects in field theory tend to expand since it is energetically favorable for them to do so \cite{Witten:1998cd,Loaiza-Brito:2001yer}, although admittedly this argument does not readily apply to the R7-brane which is a gravitational soliton, rather than a field-theoretic one. On the other hand, the appearance of a conserved charge (as captured by the bordism group) implies that whatever the ultimate fate of our object, it cannot completely disappear.

Indeed, the fact that the R7-brane could be unstable does not contradict the
observed monodromy analysis just presented since this takes place at the
boundary of the spacetime. Although we have not been able to rigorously determine the stability of  R7-branes,  many properties of the object can be studied without answering this important dynamical question. For instance, since the branes are not BPS, we expect a net attractive force between R7-branes. This can be made precise by considering
 the combined monodromy of an $\Omega$-brane and
an $\mathsf{F}_{L}$-brane. As already noted, the monodromy for this combined object is
identical to that of a supersymmetric $I_{0}^{\ast}$ singularity, i.e., a
7-brane with $\mathfrak{so}(8)$ gauge symmetry. This strongly suggests that
the $\Omega$-brane and an $\mathsf{F}_{L}$-brane are energetically attracted, forming a
supersymmetric $\mathfrak{so}(8)$ 7-brane and gravitational radiation:%
\begin{equation}
\Omega+\mathsf{F}_{L}\rightarrow\mathfrak{so}(8)+\text{radiation.}%
\end{equation}
Next, since the monodromies each square to the identity, we
conclude that pairs of the same R7-branes annihilate to radiation:%
\begin{align}
\Omega+\Omega &  \rightarrow\text{radiation}\\
\mathsf{F}_{L}+\mathsf{F}_{L}  &  \rightarrow\text{radiation}.
\end{align}
By the same token, we can also add R7-branes to $\mathfrak{so}(8)$ 7-branes to arrive at
the products:%
\begin{align}
\Omega+\mathfrak{so}(8)  &  \rightarrow \mathsf{F}_{L}+\text{radiation}\\
\mathsf{F}_{L}+\mathfrak{so}(8)  &  \rightarrow\Omega+\text{radiation}.
\end{align}

One subtlety in such recombination analyses is that there is also a
non-trivial braiding relation between the $\mathsf{F}_{L}$- and $\Omega$-brane. Indeed,
due to the low codimension, we have the rearrangement identity:%
\begin{equation}
\Omega\mathsf{F}_{L}\rightarrow \mathsf{F} \mathsf{F}_{L}\Omega,
\end{equation}
where on the righthand side, an $\mathsf{F}$ brane is associated with a spacetime fermion parity defect, which in
F-theory terms we interpret as a $dP_9$ geometry.

These annihilation processes are compatible with the braiding relations
obtained in line (\ref{Braiding}). To see why, consider the special case of
the Sen limit of F-theory on an elliptically fibered K3 surface. In this
limit, we have four $\mathfrak{so}(8)$ 7-branes, which we identify as four
separate $\Omega \mathsf{F}_{L}$ bound states. This is an on-shell \textquotedblleft
minimal energy\textquotedblright\ configuration which we can move off-shell by
rearranging the position of the individual branes. Doing so, we see that there
is a possible annihilation channel which can allow the various R7-branes to
annihilate. Observe, however, that the annihilation channel which eliminates
two $\mathfrak{so}(8)$ 7-branes simply takes us to an $\mathsf{F}$ defect, i.e., a $dP_{9}$
geometry. To see why, consider the rearrangement of a pair of $\mathfrak{so}(8)$ 7-branes:
\begin{align}
\left(  \mathfrak{so}(8)\right)  \left(  \mathfrak{so}(8)\right)   &  =\left(
\Omega \mathsf{F}_{L}\right)  \left(  \Omega \mathsf{F}_{L}\right)  \\
&  =(\Omega)(\mathsf{F}_{L}\Omega)(\mathsf{F}_{L})\\
&  =(\Omega)(\Omega \mathsf{F} \mathsf{F}_{L})( \mathsf{F}_{L})\label{joeyjoejoeshabadoo}\\
&  =(\Omega\Omega)(\mathsf{F})(\mathsf{F}_{L}\mathsf{F}_{L})
\end{align}
where in line (\ref{joeyjoejoeshabadoo}) we used the braid relation of line
(\ref{Braiding}). The pair of $\Omega$'s and $\mathsf{F}_{L}$'s can annihilate, leading
now to:%
\begin{equation}
\left(  \mathfrak{so}(8)\right)  \left(  \mathfrak{so}(8)\right)
=(\Omega\Omega)(\mathsf{F})(\mathsf{F}_{L} \mathsf{F}_{L})\rightarrow \mathsf{F} + \text{radiation,}%
\end{equation}
so the branes simply rearrange to an $\mathsf{F}$ defect (namely a $dP_{9}$ surface),
accompanied by radiation.

\subsection{Deficit Angles}

Self-consistency of the existence of the R7-brane solution also constrains the
net conical deficit angle obtained from encircling one such object. Recall
that the $\Omega$-brane and $\mathsf{F}_{L}$-brane can combine to form a supersymmetric
$\mathfrak{so}(8)$ 7-brane. The conical deficit angle around all four
$\mathfrak{so}(8)$ 7-branes leads to a trivial conical deficit angle, i.e.
$\alpha_{4\text{ }\mathfrak{so}(8)^{\prime}s}=4\pi$, so in particular, a
single $\mathfrak{so}(8)$ generates a conical deficit angle of $\alpha
_{\mathfrak{so}(8)}=\pi$. Now, since the $\Omega$-brane and $\mathsf{F}_{L}$-brane have
the same deficit angle and also combine to form an $\mathfrak{so}(8)$ stack
accompanied by radiation, we have:%
\begin{equation}
\alpha_{\Omega}+\alpha_{\mathsf{F}_{L}}=\alpha_{\mathfrak{so}(8)}+\alpha_{\text{rad}}%
\end{equation}
or:%
\begin{equation}
2\alpha_{R7}=\pi+\alpha_{\text{rad}}.
\end{equation}
Since the deficit angle generated by radiation is bounded from below, we
obtain the inequality:%
\begin{equation}\label{eq:r7deficit}
\alpha_{R7}\geq\frac{\pi}{2}.
\end{equation}
Extracting more detailed properties such as the dilaton profile is
more challenging because of strong coupling effects, a point we return to shortly.

\section{Probing R7-Branes} \label{sec:PROBE}

So far, our analysis has centered on the asymptotic properties of R7-branes
far from their core. In this section we turn to probes of R7-branes by
supersymmetric objects of type IIB\ string theory. We assume that we are working on
distance scales far from the interior of the R7-brane. If the R7 is indeed unstable, and
tends to expand in size, the probe approximation will only be
valid provided we remain outside this expanding bubble.

The probe analysis naturally splits according to the transformation rules of
the corresponding p-form potentials which couple to the worldvolume of these
branes. For stringlike excitations the reflections distinguish between F1- and
D1-string charge. For D3-branes and D7-branes, however, the R7-brane always
acts by charge conjugation on the corresponding object. Our plan in this
section will be to use these probes to extract further details of such
R7-branes.

\subsection{String Lasso}

We begin by considering stringlike probes of R7-branes. Since we know the
action of $\Omega$ and $\mathsf{F}_L$ on the two 2-form fields $B_{2}$ and
$C_{2}$ in type IIB supergravity%
\begin{equation}
\Omega:\left[
\begin{array}
[c]{c}%
C_{2}\\
B_{2}%
\end{array}
\right]  \rightarrow\left[
\begin{array}
[c]{c}%
C_{2}\\
-B_{2}%
\end{array}
\right]  ,\text{ \ \ } \mathsf{F}_L :\left[
\begin{array}
[c]{c}%
C_{2}\\
B_{2}%
\end{array}
\right]  \rightarrow\left[
\begin{array}
[c]{c}%
-C_{2}\\
B_{2}%
\end{array}
\right]  ,
\end{equation}
we can conclude that some stringlike excitations can end on the R7-brane. To
see why, consider a string with $\left[  p,q\right]  $-charge wound around the
monodromy defect. This string couples to the 2-form potential $pB_{2}+qC_{2}%
$,\footnote{In our conventions (see \cite{Weigand:2018rez}), the $SL(2,\mathbb{Z})$ matrix
$\left[
\begin{array}
[c]{cc}%
a & b\\
c & d
\end{array}
\right]  $ acts on the doublet $\left[
\begin{array}
[c]{c}%
C_{2}\\
B_{2}%
\end{array}
\right]  $ by left multiplication. The convention is chosen so that the
monodromy around a D7-brane is given by the T-transfomation $\left[
\begin{array}
[c]{cc}%
1 & 1\\
0 & 1
\end{array}
\right]  $. The charge of a $[p,q]$-string is represented by the column vector
$\left[
\begin{array}
[c]{c}%
p\\
-q
\end{array}
\right]  $. The pairing between the charge and the potential follows from the
two-index $\varepsilon$ tensor.} where $p$ tells us the number of units of F1-charge and $q$ tells us the number of units of D1-charge.

\begin{figure}[ptb]
\centering
\includegraphics[width = 0.5 \textwidth]{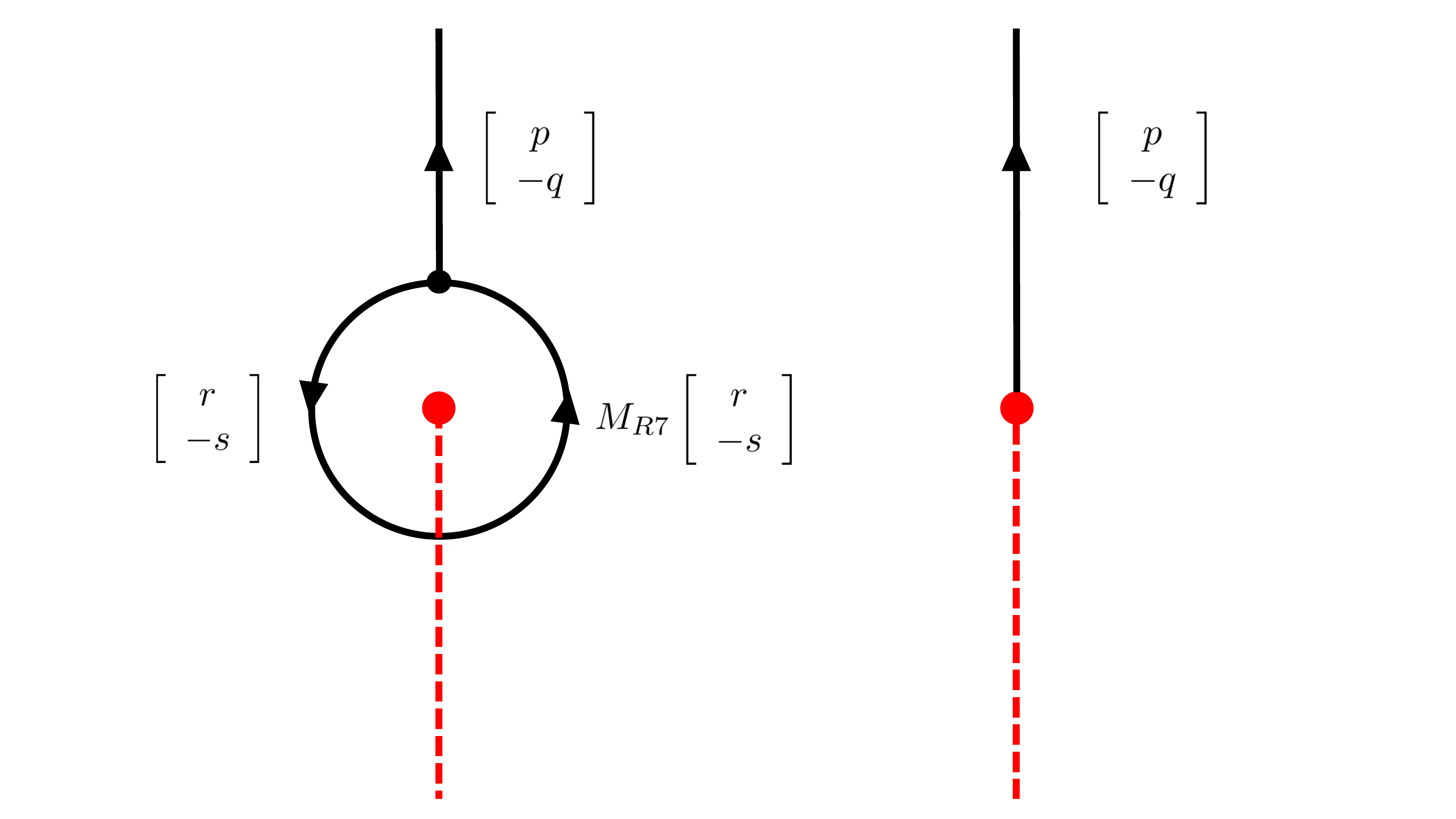}\caption{Depiction of
a lasso configuration encircling an R7-brane. There is an excess charge in the
configuration which is carried away to infinity. Shrinking the lasso to the
core of the R7-brane, we obtain a string which ends on the R7-brane. We find
that an even number of F1-strings can end on an $\Omega$-brane and an even
number of D1-strings can end on an $\mathsf{F}_{L}$-brane.}%
\label{fig:stringjunc}%
\end{figure}

We now show that such strings can end on the appropriate R7-branes. Consider \textquotedblleft lassoing\textquotedblright\ the R7-brane by surrounding it with one type of $[p,q]$ string, i.e., a loop junction.
As in \cite{Cvetic:2022uuu} (see also \cite{Cvetic:2021sxm}), the idea is to consider a $[p,q]$ string which begins at infinity and proceeds
to encircle the R7-brane. We can consider a string junction in which we consider an $[r,s]$ string exiting the junction and an $[r^{\prime},s^{\prime}]$ string entering the junction. Passing the $[r,s]$ string around the R7-brane, its charge undergoes monodromy in passing through the branch cut of the R7. We can form a closed loop precisely when the monodromy yields $[r,s] \mapsto [r^{\prime}, s^{\prime}]$. Consequently, there is an ``excess charge'' which can escape to infinity. See Figure \ref{fig:stringjunc} for a depiction of this configuration.

The excess charge carried by the asymptotic leg of the lasso configuration is then:
\begin{equation}
\left[
\begin{array}
[c]{c}%
p\\
-q
\end{array}
\right]  _{\text{asymptotic}}=(M_{R7}-\mathbf{1}_{2\times2})\left[
\begin{array}
[c]{c}%
r\\
-s
\end{array}
\right]  ,
\end{equation}
where $M_{R7}$ is the monodromy of the R7-brane in question. In the two cases
of interest, we find:\footnote{Here we use the two index epsilon tensor to present the $[p,q]$-string charge as the column vector $[p,-q]^{T}$; recall that the $SL(2,\mathbb{Z})$ invariant combination is $p B_2 + q C_2$.}
\begin{equation}
\Omega:\quad\left[
\begin{array}
[c]{c}%
p\\
-q
\end{array}
\right]  _{\text{asymptotic}}=\left[
\begin{array}
[c]{c}%
-2r\\
0
\end{array}
\right]  \,,\quad \mathsf{F}_L:\quad\left[
\begin{array}
[c]{c}%
p\\
-q
\end{array}
\right]  _{\text{asymptotic}}=\left[
\begin{array}
[c]{c}%
0\\
2s
\end{array}
\right]  .
\end{equation}
Subsequently, one contracts the loop onto the defect. The final configuration
is the asymptotic string ending on the 7-brane generating the monodromy.

We therefore see that an even number of fundamental strings can end on the
$\Omega$-brane and an even number of D-strings can end on the $\mathsf{F}_{L}$-brane. Of course, our arguments here make heavy use of the string lasso picture, and will not capture any junction that is not apparent from that perspective. For instance, it could be that a single F1 string can end on the $\Omega$-brane, but that the corresponding endpoint cannot be ``resolved'' to a configuration such as the one depicted in Figure \ref{fig:stringjunc} (see also \cite{Cvetic:2021sxm} where this was related to center 1-form symmetries of F-theory backgrounds). Although this possibility is certainly interesting, we emphasize that at least the effects described in this Section are unavoidable.

\subsection{Five-Brane Lassos}

Essentially the same argument used for $[p,q]$-strings also carries over for $[p,q]$-5-branes. Indeed, these objects are magnetic dual in ten dimensions, so the same reasoning allows us to conclude that such 5-branes can also end on an R7-brane. More precisely, an even number of D5-branes can end on an $\mathsf{F}_L$-brane, and an even number of NS5-branes can end on the $\Omega$-brane.

\subsection{Three-Brane Lasso}\label{ssec:BANALPROBE}

Consider next the case of D3-branes and 7-branes in the vicinity of the
R7-brane. Recall that for both sorts of R7-brane, the
zero-form potential and the chiral 4-form potential flip sign under
monodromy. What this means is that a probe D3-brane will, under monodromy,
convert to an anti-D3-brane, and a probe supersymmetric D7-brane will change
into an anti-7-brane. In this sense, the R7-brane functions as a
higher-dimensional generalization of an Alice string
(see \cite{Schwarz:1982ec, Schwarz:1982zt}).\footnote{An example of this sort in four dimensions is QED\ with gauged charge conjugation.}

As one moves a D3 around the R7-brane, the corresponding opposite charge must be induced in the defect, in a higher-dimensional version of the Alice string for charge conjugation discussed in \cite{Schwarz:1982ec, Schwarz:1982zt}. From this it follows that, just like ordinary D7-branes, the reflection 7-brane must have appropriate worldvolume fields that allow it to carry D3-brane charge, as we will discuss in Section \ref{ssec:WORLDVOLUME}.

\begin{figure}[t!]
\centering
\includegraphics[width = 0.6 \textwidth]{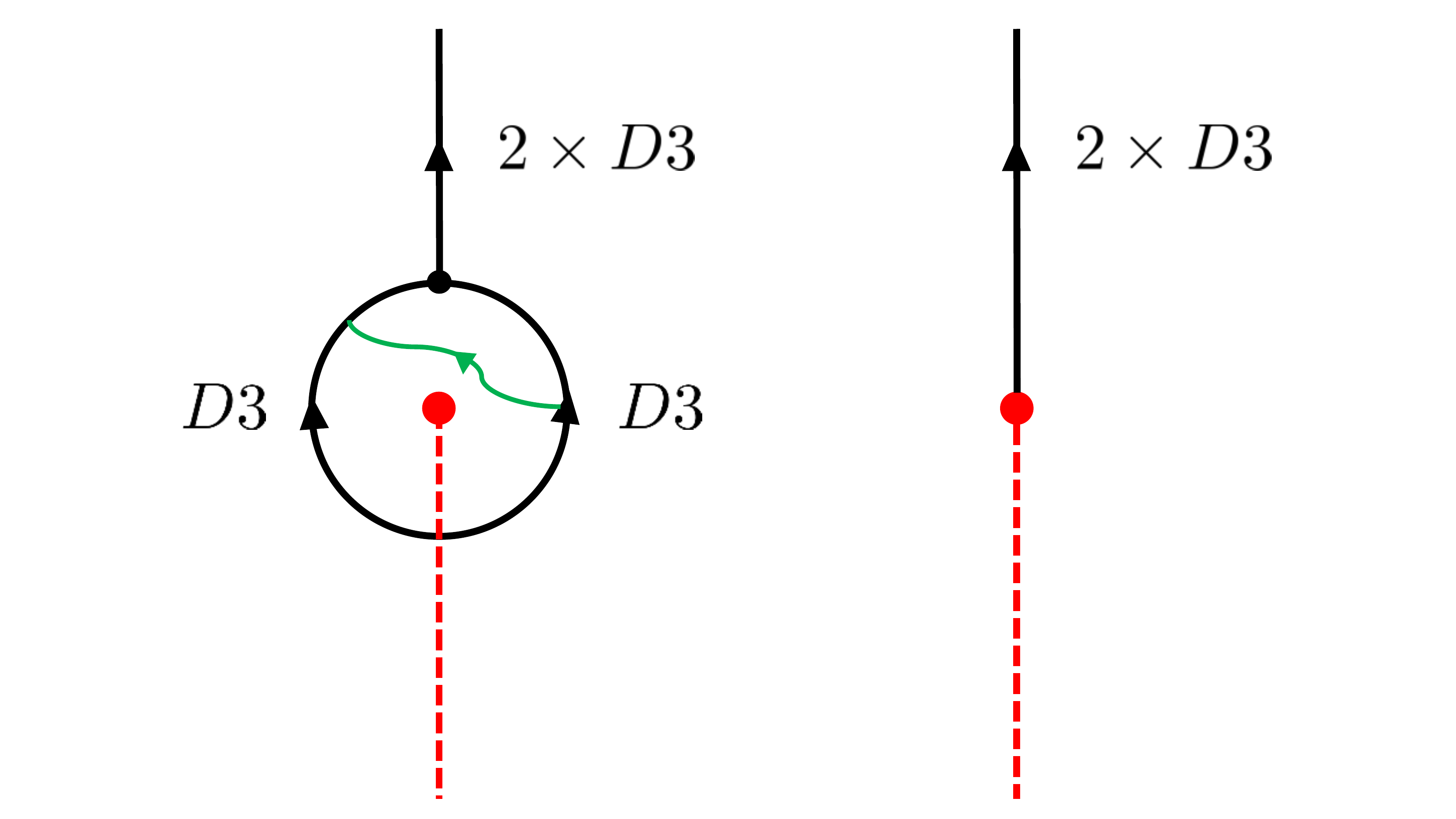}\caption{Here we illustrate a lasso configuration of D3-branes (black lines) surrounding an R7-brane (red dot). Two D3-branes extend off to infinity, and by local charge conservation, they can split off to form a junction. Since the R7-brane sends D3's to anti-D3's, the orientation of charge flow in the junction can reverse, allowing us to close off a ``lasso'' in passing through the branch cut. It is energetically favorable for this lasso to shrink to small size, leading to a pair of D3-branes which can end on the R7-brane. Similar considerations also hold for $[p,q]$ 7-brane configurations. }
\label{fig:d3lasso}%
\end{figure}

Perhaps more surprisingly, we can also form a D3-brane lasso around the R7-brane. Indeed, because a D3-brane converts to an anti-D3 in encircling the R7-brane, we can build a lasso configuration of the sort depicted in figure \ref{fig:d3lasso}, namely a pair of D3's extend to infinity, and splits off into a junction of individual D3-branes. We can join up these D3-branes by passing them around the branch cut of the R7-brane, since in doing so, D3-brane charge is conjugated. Since parallel D3-branes form a BPS configuration, there is an energetic preference for this loop to shrink to zero size, and in this limit, we observe a pair of D3-branes ending on the R7-brane. We comment that this does not happen for ordinary $[p,q]$ 7-branes, since the monodromy of a D3-brane around that configuration is trivial.

\subsection{7-Brane Probes}

Finally, let us turn to the case of 7-brane probes. Moving a D7-brane around the R7-brane has a stronger backreaction,
and the Alice effect is instead encoded in the monodromy action on 7-branes. Given an element of $SL(2,\mathbb{Z})$, we can conjugate by either the $\Omega$-brane or the $\mathsf{F}_L$-brane to extract the monodromy. For example, we have:
\begin{equation}
\left[
\begin{array}
[c]{cc}%
a & b\\
c & d
\end{array}
\right]  \overset{R7}{\mapsto}\left[
\begin{array}
[c]{cc}%
a & -b\\
-c & d
\end{array}
\right] ,
\end{equation}
for both sorts of R7-branes. In particular, for a D7-brane this produces an anti-D7-brane. More generally, if we take a $[p,q]$-7-brane with monodromy matrix:
\begin{equation}
M_{[p,q]}=\left[
\begin{array}
[c]{cc}%
1+pq & p^{2}\\
-q^{2} & 1-pq
\end{array}
\right],
\end{equation}
around an R7-brane with monodromy given by
\begin{align}
M^{-1}_{[p,q]} \, M_{R7} \, M_{[p,q]} \,,
\end{align}
one obtains the associated anti-$[p,q]$-7-brane.

The lasso configuration for 7-branes is more complicated due to the branch cuts in the configuration and we leave a study of these setups for future work.

\section{R7-Brane Worldvolume} \label{sec:ANOMO}

In this section we study the worldvolume theory of R7-branes. In particular, we present
evidence that they have gapless degrees of
freedom in their worldvolume theory, as well as non-trivial topological sectors. While this suggestively hints at the
possibility of an interacting fixed point in $D>6$ spacetime dimensions, we
defer a more complete analysis of this tantalizing possibility to future work.

To establish these properties, we will first argue that R7-branes can be
viewed as end of the world branes in well-known perturbative 9D\ string
backgrounds. With this in place, the task of determining the anomaly
polynomial for R7-branes amounts to computing the relevant anomalies in the
bulk of the 9D\ model. The mechanism of anomaly inflow \cite{Callan:1984sa} then
allows us to deduce the anomalies on the boundary R7-branes. We show that
the anomalies are cancelled if the 8D worldvolume has a massless spectrum containing
a Weyl fermion, or something anomaly-equivalent to it. In case it was an 8D weyl fermion, it would naturally serve as the goldstino associated to breaking 8D $\mathcal{N}=1$ supersymmetry.
Additionally, in order for the BPS branes of type IIB to consistently end on the R7,
we argue that topological degrees of freedom on the R7 can indeed
be inferred from conservation of RR charge.

The two perturbative string backgrounds of interest are the asymmetric
orbifold of type IIB\ (AOB), and its S-dual counterpart the Dabholkar-Park
(DP) background. The $\mathsf{F}_{L}$-brane serves as an end of the world (EOW)\ brane
for the AOB\ background, and the $\Omega$-brane serves as an EOW\ brane for
the DP\ background. Since the two configurations are S-dual, it is enough to
focus on the AOB\ background. For further discussion of the various
backgrounds, as well as closely related 9D\ vacua, see \cite{Aharony:2007du}.

The rest of this section is organized as follows. We begin with a brief review
of the\ AOB\ and DP\ backgrounds.  With these results in
place, we next extract the (perturbative)\ anomaly polynomial for the
8D\ theory specified by the worldvolume of the $\mathsf{F}_{L}$-brane. In Appendix \ref{app:CS1loop} we
provide additional details on the structure of anomalies as obtained from circle reduction.

\subsection{AOB and DP Backgrounds}
\label{subsec:AOBDP}

Let us now briefly review the different backgrounds of interest. We shall be
interested in compactifications of type IIB\ strings on a circle of length
$2\pi R_{9}$. The type IIB\ asymmetric orbifold (AOB)\ background is obtained
by gauging the $\mathbb{Z}_{2}$ symmetry
generated by $H_{9} \mathsf{F}_L$ whose action is:
\begin{equation}
\text{AOB: } (X^{9}\rightarrow X^{9}+\pi R_{9})\times \mathsf{F}_L.
\end{equation}
The S-dual of this system is the Dabholkar-Park background, as obtained by
gauging the $\mathbb{Z}_{2}$ symmetry generated by $H_{9}\Omega$:%
\begin{equation}\label{eq:DPAOBactions}
\text{DP: }\left\{
\begin{array}
[c]{c}%
X^{\mu}(u,\overline{u})\rightarrow X^{\mu}(\overline{u},u)\text{ \ \ for
\ \ }\mu\neq9\\
X^{9}(u,\overline{u})\rightarrow X^{9}(\overline{u},u)+\pi R_{9}\text{
\ \ \ \ \ \ }%
\end{array}
\right\}  .
\end{equation}

From the target space viewpoint these backgrounds are obtained from 10D\ type IIB\ strings compactified
on an $S^{1}$ with a suitable Wilson line switched on along the circle. The low energy effective theory consists of a 9D $\mathcal{N}=1$ supergravity multiplet and $U(1)$ vector multiplet which enhances to an $SU(2)$ vector multiplet at a self-dual radius in the AOB case.\footnote{This is due to the fact that the AOB background is self-dual under T-duality. For a review of the various components of the 9D $\mathcal{N}=1$ rank-1 moduli space of string/F-/M-theory see \cite{Aharony:2007du, Montero:2022vva}.} An R7-brane serves as an EOW brane for these 9D theories since we can interpret this configuration as type IIB\ compactified on a long
cylinder with certain Wilson lines, with the R7-brane placed at the end of the cylinder. This is similar to the case of IIB on a long cylinder, without any Wilson lines, ending on two stacks each consisting of four D7-branes and one O7$^-$ brane (or $I^*_0$ singularities in F-theory language) placed at $X^9=0$ and $\pi R_9$.

Finally, we mention that one can also consider the discrete theta angles for AOB and DP backgrounds recently found in \cite{Montero:2022vva}. These result from a non-trivial value of $\int_{S^1} C_0\in \{0,1/2\}$, much like the Sethi string \cite{Sethi:2013hra}, except they have physical consequences on the spectrum since they cannot be undone by chiral rotations of fermions. The AOB and DP string theories have a non-BPS D7-brane with K-theory charge $\mathbb{Z}_2$ and this acts as a domain wall between the AOB/DP backgrounds with $\int_{S^1} C_0=1/2$ and $\int_{S^1} C_0=0$. Therefore, an EOW brane for the AOB/DP background with a non-trivial theta angle is a fusion between the non-BPS D7-brane and an R7-brane. Turning on the $\theta$ angle can also be regarded as changing the holonomy around the $S^1$ from $\Omega$ or $(-1)^{F_L}$ to $ T(-1)^{F_L}$ or $T\Omega$, where $T$ is just the $T$ generator of $SL(2,\mathbb{Z})$, which corresponds to the second of the two $\mathbb{Z}_2$ factors of \eq{mhht}.

\subsection{Symmetric Mass Generation}
From the point of view of the AOB and DP backgrounds, the interpretation of the R7-branes is simple: they provide a boundary condition for the theory, similar to a bubble of nothing \cite{Witten:1981gj} or the $E_8$ wall in M-theory \cite{Horava:1996ma}. Indeed, one formulation of the Cobordism Conjecture is the requirement that every consistent quantum gravity admits a boundary condition. The reflection 7-brane can be obtained by demanding that this happens for either the AOB or DP backgrounds. Applying the same argument to the AOB/DP backgrounds with discrete theta angle turned on found in \cite{Montero:2022vva} merely yields a reflection 7-brane with monodromy $\Omega T$, see also the last paragraph of the previous section. Hence, we will not discuss this last example.

This perspective can be used to argue that the R7-brane should be strongly coupled, and that the dilaton should flow to a large value at its core. The argument is similar to the ones that can be made for the IIA/IIB domain wall, and is essentially that, since anomalies cancel by a Green-Schwarz mechanism, a boundary condition cannot be constructed at weak coupling. Let us review this in detail. First, consider the theory of a single 9D dilatino, and suppose that we want to construct a boundary condition for it. The modern understanding of the 9D theory of a single Majorana fermion in a manifold $X_9$ is as a boundary mode for a topological field theory in ten dimensions, which in this case is simply the  $\eta$ invariant of a ten-dimensional Majorana-Dirac operator. The relevant bordism group
\begin{equation}
\Omega_{10}^{\text{Spin}}= \mathbb{Z}_2 \times \mathbb{Z}_2 \times \mathbb{Z}_2
\end{equation}
is non-vanishing on one of the classes (represented by an exotic 10-sphere \cite{Witten:1982fp}). This not only means that there is a global gravitational anomaly and one cannot make sense of the dilatino theory by itself; it also means that there is no boundary condition for the dilatino field in 9D. A boundary condition can be regarded as a domain wall between a phase where the dilatino is massless and another one where the dilatino is gapped; but there is no symmetry-preserving gapped boundary condition  for any TQFT with non-vanishing value on a sphere \cite{Cordova:2019bsd}. Therefore, there can be no boundary condition for the dilatino by itself. As a cross-check of this, notice that any fermion that admits a mass term also admits a boundary condition, since one can just take a space-dependent mass from zero to a very large value. But there is no mass term for a 9D dilatino, as the requisite mass matrix is symmetric rather than antisymmetric \cite{Witten:2015aba,ortin2004gravity}.

The anomaly of the dilatino is exactly cancelled by that of the gravitino, but this in turn has another anomaly \cite{Lee:2022spd} on $\mathbb{HP}^2\times T^2$. Similar arguments show that one cannot find a single boundary condition for the combined gravitino and dilatino system. As explained in \cite{Lee:2022spd}, however, these anomalies can be cancelled by discrete fields, or topological couplings. Once these discrete fields are taken into account, the anomaly cancels, and it is possible to construct a boundary condition in principle. Such a boundary condition will mix the gravitino, dilatino, and discrete fields; and it cannot happen at weak coupling, simply because there is no Lagrangian term that one can write down that would gap these fields at weak coupling. Nevertheless, it can happen in principle at strong coupling, an example of the phenomenon dubbed as symmetric mass generation in \cite{Wang:2022ucy}. For this to be the case, the core of the R7-brane should be strongly coupled. We will not study this mechanism in detail, since a full treatment would require additional knowledge of the R7-brane worldvolume; we simply note that the same must be true of other domain walls predicted by the Cobordism Conjecture, such as the domain wall between IIA and IIB.

There is a further refinement of the above story that also involves the topological terms that will be discussed in the next section. As we have established above, the AOB or DP backgrounds have a gauged $SO(2)$ symmetry coming from the isometry in the compactification space. Although the nine-dimensional supercharge is uncharged under this symmetry, since the 9D fields are all zero modes from the 10D perspective, there is a closely related symmetry that affects the supercharge. The dimensional reduction on a circle of any theory with Lorentz/Euclidean invariance has a reflection symmetry. To see why, consider $x$ to be a coordinate of the non-compact space, and let $\theta$ parametrize the internal $S^1$. Reflecting $x\rightarrow -x$ or $\theta\rightarrow-\theta$ may or may not be a symmetry, depending on the properties of the parent theory. However, the combined action of two reflections
\begin{equation}
(\theta,x)\rightarrow (-\theta,-x)\label{e2344}
\end{equation}
is an isometry of determinant $+1$, and as such, is part of the Lorentz group in the higher-dimensional theory. Since this isometry is preserved by the circle background, it remains as a symmetry of the lower-dimensional theory. From the nine-dimensional point of view, where we forget about $\theta$ and keep only $x$, it is a reflection. Furthermore, it squares to $-1$ on fermions, since \eq{e2344} uplifts to a rotation by $\pi$ in the parent theory, whose square is $(-1)^F$. As a result, the reduced theory has what we call a $\mathsf{Pin}^-$ structure \cite{Witten:2015aba}. This applies to the AOB and DP backgrounds; indeed, the only reflection symmetry compatible with 9D $\mathcal{N}=1$ supersymmetry is precisely $\mathsf{Pin}^-$ \cite{Montero:2020icj}.

If the theory has a $\mathsf{Pin}^-$ symmetry, the anomalies are now classified by the much larger group
\begin{equation}
\Omega^{\mathsf{Pin}^-}_{10}=\mathbb{Z}_2 \times \mathbb{Z}_8 \times \mathbb{Z}_{128}.
\end{equation}
In principle, the anomalies should cancel here possibly against discrete terms such as in \cite{Lee:2022spd}, plus an additional contribution of potential Green-Schwarz terms in the action. For instance, a term such as:
\begin{equation}
\int A\wedge X_8, \label{term8}
\end{equation}
where $X_8$ is a combination of gravitational classes, can contribute to a Pin$^-$ anomaly, since the vector fields $A$ are odd under the $\mathsf{Pin}^-$ symmetry \cite{Montero:2020icj}. Again, a full study of anomaly cancellation in the AOB and DP backgrounds would be very interesting but lies beyond the scope of this work. The only point that we wish to make is that, because \eq{term8} participates in anomaly cancellation together with the dilatino and gravitino, a boundary condition must mix all three, and this cannot possibly happen at weak coupling.

\subsection{Worldvolume Degrees of Freedom}\label{ssec:WORLDVOLUME}

Having just argued that the R7-brane is strongly coupled, one might wonder whether we can say anything about its worldvolume theory. One fruitful technique to learn about the worldvolume theories of strongly coupled objects is anomaly inflow; see \cite{Martucci:2022krl} for a similar, recent application in the context of F-theory for low codimension objects (as is the case here). The analysis of Appendix \ref{app:CS1loop} establishes that there is an anomaly inflow from the bulk of an AOB / DP background to the respective $\mathsf{F}_L$ / $\Omega$-brane, and this can be cancelled by the anomaly of a single Weyl fermion, charged under the $U(1)_R$ symmetry which corresponds to rotations of the normal bundle, or something anomaly-equivalent to it.

Another interesting way to obtain information about the worldvolume degrees of freedom is using the lasso arguments of section \ref{sec:PROBE}. Consider, for concreteness, the AOB background. As shown for the strings in Figure \ref{fig:stringjunc}, one expects (at least even numbers of) D1-, D3-, and D5-branes to be able to end on the R7-brane. This phenomenon also happens for ordinary D-branes \cite{Johnson:2003glb}, and famously, this requires the presence of worldvolume degrees of freedom to absorb the charge of the brane. For instance, consider say a D5-brane ending on a D7-brane. The D5-brane has a worldvolume coupling,
\begin{equation} \int_{D5} C_6 \,, \end{equation}
which is not gauge invariant under the gauge transformation $C_6\rightarrow C_6+d\lambda_5$ if the D5 worldvolume has a boundary. The variation is precisely
\begin{equation}  \int_{\partial D5} \lambda_5 \,,\end{equation}
and since the boundary of the D5 is contained on the D7, it can be cancelled by a coupling which also transforms under $C_6\rightarrow C_6+d\lambda_5$ as\footnote{$\text{PD}$ denotes the Poincar\'e dual of the boundary of the D5-brane $\partial D5$ in the D7-brane worldvolume.}
\begin{equation}  -\int_{D7} \lambda_5\wedge \text{PD}\,(\partial D5) \,.\end{equation}
In the D7 worldvolume theory, this happens because there is a localized $U(1)$ gauge field, with field strength $f_2$, and Chern-Simons coupling
\begin{equation} \int_{D7} C_6\wedge f_2 \,,\end{equation}
whose gauge variation equals, upon integration by parts,
\begin{equation} \int_{D7} \lambda_5 \wedge df_2 \,.\end{equation}
Hence, if we identify $df_2$ with $ \text{PD}(\partial D5)$, that is, if the endpoint of a D5-brane on a D7 is a magnetic monopole for the worldvolume $U(1)$ gauge field, the whole system will be gauge invariant.

The same argument works for the $\mathsf{F}_L$-brane, leading to the conclusion that it must have a localized $U(1)$ gauge field with field strength $f_2$ and worldvolume coupling
\begin{equation} \int_{R7} C_6\wedge f_2 \,.\end{equation}
Therefore, the $U(1)$ gauge field can account for D5-branes ending on the $\mathsf{F}_L$-brane. Notice that the $C_6$ field is odd under $(-1)^{F_L}$, and therefore the $U(1)$ field strength $f_2$ must also be odd. Similar considerations apply for D1-branes ending on the $\mathsf{F}_L$-brane.

We must also take into account the fact that D3's can end on the R7-brane as well. To do this in the same spirit as above, we need a coupling
\begin{equation} \int_{R7} C_4\wedge X_4 \,,\end{equation}
where $X_4$ is some R7 worldvolume class to be determined. One natural option is to take $X_4 = f_2 \wedge f_2$, but this has two problems: on the one hand, its integral vanishes on $S^4$, while we would need $\int_{S^4} X_4\neq 0$ to capture the ingoing D3-brane charge. On the other hand, $C_4$ is odd under $(-1)^{F_L}$, and thus $X_4$ must be odd as well. Since it is quadratic in the field strength, it is impossible for $f_2 \wedge f_2$ to be odd. Both these obstacles are remedied if we postulate that
\begin{equation} X_4=da_3 \,,\end{equation}
where $a_3$ is a \emph{massless 3-form gauge field} living on the worldvolume of the R7-brane. Although we have not proven this is the only kind of structure that can account for the D3-brane charge inflow required by the lasso arguments in Section \ref{sec:PROBE}, it is certainly the simplest solution. As emphasized above, $a_3$ is odd under reflections, meaning that its field strength takes values in a cohomology group with twisted coefficients.\footnote{The coefficients are given by a local system as specified by the appearance of a non-trivial reflection bundle. The analogous feature is already present in the 3-form potential of 11D supergravity \cite{Witten:1996md,Witten:2016cio}.}

To the best of our knowledge, this is the first appearance of a massless 3-form as a worldvolume field of a brane in string theory, as opposed to a spacetime field. The existence of the massless 3-form raises tantalizing questions: for instance, the electrically charged objects coupled to a 3-form are membranes, and  although the codimension of the object is very low, so that gravitational backreaction cannot be neglected in principle, one can speculate about the existence of a ``little membrane'' description of the theory at high energies.

We cannot resist presenting another speculation before closing this section. We have seen that the R7-brane breaks supersymmetry, has interesting worldvolume dynamics, and is a strongly coupled object. All these could be taken as hints that the low-energy dynamics of the R7-brane corresponds to a non-supersymmetric CFT in eight dimensions. Although we certainly do not have an argument that this is the case, we believe that the idea certainly warrants attention, and we hope to come back to it in the future. A first example of a non-supersymmetric CFT in high dimensions would have far reaching implications in both quantum gravity and quantum field theory.

\section{Doubled F-theory} \label{sec:DOUBLE}


From a practical point of view, one of the key features of F-theory is that it
provides a geometric characterization of non-perturbative type IIB vacua in
terms of elliptically fibered Calabi-Yau spaces (or more generally, genus one
fibered spaces). Indeed, the existence of the elliptic fibration provides a
convenient way to encode an $SL(2,\mathbb{Z})$ duality bundle on the
10D\ spacetime. For Calabi-Yau spaces, the existence of a Killing spinor is
guaranteed, and thus implies that the Spin structure of the Calabi-Yau induces a
Spin-$Mp(2,\mathbb{Z})$ duality bundle. An important feature of this
formulation is that the Ricci-flatness condition in the total space of the fibration also implies that we solve
the Einstein field equations in the presence of a possibly non-trivial
axio-dilaton profile. This in particular allows us to read off intricate
configurations of intersecting 7-branes in type IIB\ vacua.

Given the fact that the full IIB\ duality group is actually $GL^{+}%
(2,\mathbb{Z})$, it is natural to ask whether we can develop a similar
formulation for more general backgrounds. As a preliminary step in this
direction, we will first discuss the $GL(2,\mathbb{Z})$ bundle detected by bosons. Note that because $GL(2,\mathbb{Z})$ has determinant
$-1$ elements, a principal $GL(2,\mathbb{Z})$ bundle will not be captured by
an elliptic curve, simply because we must allow for orientation reversal on
the torus. Since any non-orientable space has an oriented double cover, one
approach to building geometries with a $GL(2,\mathbb{Z})$ bundle is to work in the double cover, which has an
$SL(2,\mathbb{Z})$ bundle and can therefore be described by an ordinary genus one fibration.

Now, whereas F-theory on an elliptically fibered Calabi-Yau space
automatically furnishes us with a solution to the type IIB\ equations of
motion, this need not be the case for these more general doubled
configurations. Indeed, these configurations are helpful in characterizing
possible \textquotedblleft off-shell\textquotedblright\ brane configurations
which will ultimately relax to \textquotedblleft
lower-energy\textquotedblright\ equilibrium configurations. It is nevertheless
useful to specify these duality bundles geometrically because it provides a
concise way to encode the monodromy of all the codimension two defects in a
way which is consistent with constraints such as tadpole cancellation. Just as
in standard F-theory backgrounds, a doubled F-theory background is
supersymmetric whenever the corresponding dual M-theory background is supersymmetric.

To encode reflection 7-branes and possibly more general Spin-$GL^{+}%
(2,\mathbb{Z})$ duality bundle structures, we can build examples by specifying
a pair $(\widehat{X},\rho)$, where $\widehat{X}\rightarrow\widehat{\mathcal{B}%
}$ is a genus one fibration and $\rho$ is an involution which preserves the
orientation of the base $\widehat{\mathcal{B}}$. The restriction that $\rho$
preserves the base orientation is necessary since IIB string theory is not invariant under
reflections. The case where $\rho$ is orientation-preserving can always be
reduced to trivial $\rho$, because the quotient by $\rho$ also produces a
genus one fibration. So, the only interesting case is when $\rho$ is
orientation-reversing along the fibers. The action of $\rho$ on fermions should be of $\mathsf{Pin}^+$ type as required by M- / F-theory duality.

In this section we will focus on some examples of local models with R7-branes.
It would be quite interesting to build more general
compact models in doubled F-theory (\reflectbox{F}F-theory) with such structures,
but we leave this to future work.

\subsection{Geometry of a Single R7-Brane}

Let us see how the framework of doubled F-theory captures reflection 7-branes.
We first consider a trivial $T^{2}$ fibration over a circle, parametrized as
$(\theta,z)$ with the identifications
\begin{equation}
\theta\sim\theta+2\pi,\quad z\sim z+1\sim z+\tau
\end{equation}
and quotient by the orientation reversing (free) action
\begin{equation}
\rho_{\mathsf{F}_{L}}:\,(\theta,z)\,\rightarrow\,(\theta+\pi,\overline{z}).
\end{equation}
For this to be a symmetry of the lattice, one must require that $\tau$ is
purely imaginary.\footnote{Again, we neglect the more general possibility that $\tau=1/2+\frac{i}{g_s}$ which is relevant for the backgrounds studied in \cite{Montero:2022vva}.} The pair $(T^{3},\rho)$ thus constructed is the doubled
F-theory description of a IIB compactification with a Wilson line of
$(-1)^{F_{L}}$ IIB symmetry, precisely the AOB background described in Section \ref{subsec:AOBDP}, see \cite{Hellerman:2005ja,Aharony:2007du}, the F-theory description of this background is explained in \cite{Montero:2022vva}.
Unlike similar constructions with the holonomy
of ordinary 7-branes, this background preserves sixteen supercharges.

A similar construction, using the orientation reversing action
\begin{equation}
\rho_{\Omega}:\,(\theta,z)\,\rightarrow\,(\theta+\pi,-\overline{z})
\end{equation}
yields a IIB compactification with a Wilson line of $\Omega$, the Dabholkar-Park background of Section \ref{subsec:AOBDP}.
As described there, this is another supersymmetric compactification, S-dual to the AOB background. In both cases, the action on the torus lattice
parameter $\tau$ maps it to $-\bar{\tau}$, which forces the real part to be 0 or $\pi$.
We will mostly set $\mathrm{Re} \, \tau=0$, but our construction can equally accommodate the $\mathrm{Re} \, \tau=\pi$ vacua which have only recently been explored \cite{Montero:2022vva}.

Our aim will be to \textquotedblleft fill in\textquotedblright\ the circle, to
describe the cobordism defect at the core of the AOB and DP background we have
just constructed. To proceed further, let us return to the \textquotedblleft
trivial\textquotedblright\ Weierstrass model:%
\begin{equation}
y^{2}=x^{3}+f_{0}x+g_{0},
\end{equation}
where here, $f_{0}$ and $g_{0}$ are constant. As explained in Appendix \ref{app:weierstrass},
the two reflections act on the holomorphic coordinates $x$ and $y$ via:%
\begin{equation}
R_{\mathsf{F}_{L}}:(x,y)\,\rightarrow\,(\overline{x},\overline{y})\quad\text{and}\quad
R_{\Omega}:(x,y)\,\rightarrow\,(\overline{x},-\overline{y}).
\end{equation}
We observe that the Weierstrass equation is invariant under this action
provided we take $f$ and $g$ to be purely real.

The doubled F-theory description of an R7-brane is then simply
\begin{equation}\label{eq:doubleweierstrass}
y^{2}=x^{3}+f_{0}x+g_{0},\quad f_{0},g_{0}\,\in\mathbb{R},\quad(z,x,y)\,\sim
\,(-z,\overline{x},\pm\overline{y}).
\end{equation}
Away from $z=0$, the action is free. At $z=0$ we have a singularity, which is
where the R7-brane is located. In the description of ordinary 7-branes, we are
used to the torus fiber pinching at loci of the base. In the doubled F-theory picture, one might then be tempted to conclude that the fiber just degenerates in a simple way, for instance, as an interval or a Mobius strip. While this makes sense at the level of the geometry, there is an issue with defining spinors. The Spin structure along the circle with coordinate $\mathrm{arg}(z)$ in the double cover, $\widehat{X}$ must have periodic Spin structure in order for
$\widehat{X}/\rho$ to have $\mathsf{Pin}^+$ structure. This is because the torus formed from the $\mathrm{arg}(z)$ coordinate, and either the $a-$ or $b-$ cycle of the torus fiber of $\widehat{X}$ (depending on if the R7 is a $\mathsf{F}_L$ or $\Omega$ brane), becomes a Klein bottle after quotienting by $\rho$. We then see from Appendix \ref{app:kb} that the base circle of a Klein bottle must have periodic Spin structure in the torus double cover in order for the Klein bottle to have $\mathrm{Pin}^+$ structure.  This means that, to define the R7-brane, we must choose a periodic spin structure in the double cover, i.e. we must ``remove the point'' where the R7-brane is located, which introduces a new 1-cycle along which fermions are periodic. A natural possibility then is that presence of the R7 causes the base of $\widehat{X}$ to be asymptotic to a cylinder rather than a cone with deficit angle $\alpha<2\pi$. In the physical spacetime, $\widehat{X}/\rho$, this translates to the deficit angle of the R7 satisfying $ \alpha_{R7}\geq \pi$. While this is consistent with the lower bound mentioned in \eqref{eq:r7deficit}, we do not have an analytic handle on the (presumably time-dependent) metric produced by the R7-brane so we cannot yet say if this minimal possibility is correct.

\subsection{Including Other 7-Branes}

We also mention how one can write double F-theory
fibrations with R7-branes in the presence of other 7-branes. We still work in a local patch, but the
complex structure of the double cover torus can now be fibered non-trivially over
the spacetime. To present an explicit example, consider the doubled F-theory
model:
\begin{equation}
y^{2}=x^{3}+(z-a)^{2}(\overline{z}+\overline{a})^{2}\,x+\alpha(z-a)^{3}%
(\overline{z}+\overline{a})^{3},\quad(z,x,y)\,\sim\,(-z,\overline{x}%
,\overline{y}).
\end{equation}
for $a\neq0$. Close to the origin, where there is a fixed point, the model
describes the $\mathsf{F}_{L}$-brane. Close to $z=\pm a$, the model describes a local
$I_{0}^{\ast}$ singularity. The model as a whole then describes the bound
state of a $\mathsf{F}_{L}$-brane and an $\mathfrak{so}(8)$ 7-brane. As expected from
the analysis of monodromies already given, the total monodromy at infinity is
that corresponding to an $\Omega$-brane.

\section{Conclusions} \label{sec:CONC}

In this paper we have used the Cobordism Conjecture, in tandem with
the known duality symmetry of IIB\ strings / F-theory to predict the existence
of a new class of reflection 7-branes, which we dubbed R7-branes.  Such objects are
predicted to exist because the bordism group $\Omega_{1}^{\mathcal{G}}$ for
$\mathcal{G}=$ Spin-$GL^{+}(2,\mathbb{Z})$ is non-trivial. Under monodromy around such an
object, there is a corresponding reflection on the torus of F-theory. There
are two possible reflections, which correspond to an $\Omega$-brane and an
$\mathsf{F}_{L}$-brane. We have shown that these branes break all supersymmetries, that
the core of these objects exists at order one values of the axio-dilaton, and
that these objects are potentially unstable. Our approach to studying these
objects has involved first determining some basic features of monodromy,
proceeding next to various brane probes, and finally to a study, via anomaly
inflow, of the degrees of freedom localized on such objects. We put
forward a proposal for \textquotedblleft doubled F-theory\textquotedblright%
\ (\reflectbox{F}F-theory) configurations which can be used to characterize the topological features of
such objects. In the remainder of this section we discuss some avenues for
future investigation.

We have argued that the axio-dilaton likely approaches an order one value in
the core of the R7-branes. It
would be most instructive to find a more microscopic characterization of these strong coupling features,
or alternatively, to find related weakly coupled avatars
which might be used to build closely related analogs of these same systems.

One of the basic features of F-theory with reflections is that our main
examples have all been non-supersymmetric. It would be interesting to
determine backgrounds where supersymmetry is preserved, but in which there is
no holomorphic Calabi-Yau geometry present. We find it suggestive that
similar non-holomorphic structures have been considered in the context of
F-theory on $Spin(7)$ backgrounds (see \cite{Bonetti:2013fma, Bonetti:2013nka,
Heckman:2018mxl, Heckman:2019dsj, Cvetic:2020piw, Cvetic:2021maf}).

Perhaps even more intriguing would be to use these non-supersymmetric
R7-branes as an ingredient in the construction of non-supersymmetric vacua.
Quite likely, the best we can hope to achieve in this setting is a
sufficiently long-lived metastable configuration of such branes. This holds
out the possibility of building new classes of string-based models where
topological considerations can serve as a strong constraint on the ultimate
fate of these constructions.

Our anomaly inflow analysis strongly suggests that the R7-branes support
gapless degrees of freedom. Since we are dealing with an intrinsically
non-supersymmetric system at strong string coupling, it is tempting to speculate that this construction
realizes a non-supersymmetric conformal fixed point in eight dimensions, which
is of course higher than the limit allowed by supersymmetry \cite{Nahm:1977tg}.
To make this system stable, we would also need to take a limit in which the string
coupling is tuned to zero, much as one takes a decoupling limit for coincident
M5-branes to reach the celebrated $\mathcal{N}=(2,0)$ superconformal field theories in
six dimensions. To truly demonstrate that we have an interacting fixed point,
i.e., that we have non-trivial 3-point functions will likely require
further detailed computations such as the calculation of various graybody
factors on such branes (see e.g., \cite{Gubser:1996de, Gubser:1997yh}).
We leave an investigation of such possibilities for future work.

\section*{Acknowledgments}

We thank H.Y Zhang for collaboration at an early stage of this work. We thank
A. Debray for several helpful discussions and collaboration on related work. We thank
S. Goette, A. Grassi, M. H\"ubner, A.P. Turner, C. Vafa, and X. Yu for helpful discussions. MD thanks Harvard University and the CMSA for its hospitality during the final stages of this manuscript. MM thanks UPenn for its hospitality and the stimulating atmosphere in the early stages of this project. MD and MM gratefully acknowledge the ``Engineering in the Landscape: Geometry, Symmetries and Anomalies'' workshop in Uppsala University for providing a stimulating research environment in which parts of this work were advanced. JJH, MM and ET thank the Simons Summer Workshop 2022 for hospitality while this work was being completed.
Part of this work was performed at the conference “Geometrization of (S)QFTs
in $D \leq 6$” held at the Aspen Center for Physics, which is supported by National Science
Foundation grant PHY-1607611. JJH also acknowledges the fine people at Arby's for consistently providing an outstanding selection of roast beef based products \cite{Arbys}. The work of MD is supported by the German-Israeli Project Cooperation (DIP) on ``Holography and the Swampland''.
The work of JJH and ET is supported by DOE (HEP) Award DE-SC0013528.
The work of MM is supported by a grant from the Simons Foundation (602883, CV)
and by the NSF grant PHY-2013858.

\appendix

\section{Elliptic Curves and Weierstrass $\wp$-Functions}\label{app:weierstrass}

Recall that for an elliptic curve with complex structure $\tau$
the Weierstrass $\wp$-function is defined as:%
\begin{equation}
\wp(z;\tau)=\frac{1}{z^{2}}+\underset{(m,n)\in\mathbb{Z}^2\neq (0,0)
}{\sum}\left(  \frac{1}{(z+m+n\tau)^{2}}-\frac{1}{m+n\tau%
}\right)  ,
\end{equation}
which satisfies the following differential equation
\begin{equation}\label{eq:weierstrassmodel}
  \left( \wp(z;\tau)^\prime\right)^2= \wp(z;\tau)^3+f(\tau) \wp(z;\tau) +g(\tau)
\end{equation}
provided the definitions
\begin{equation}\label{eq:fandgweierstrass}
f=-4^{1/3}\cdot60\sum_{(m,n)\neq(0,0)}\frac{1}{(m+n\tau)^{4}},\quad
g=-540\cdot60\sum_{(m,n)\neq(0,0)}\frac{1}{(m+n\tau)^{6}}.
\end{equation}

Now that we have the Weierstrass coefficients $f$ and $g$ written as modular forms of $\tau$, we see that the requirement that $\tau$ is purely imaginary restricts the $f$
and $g$ functions to be real themselves. From \eqref{eq:fandgweierstrass},
we also see that $f<0$, $g>0$.

To understand the action of the orientation-reversing isometries described
in Section \ref{sec:DOUBLE}, we can use the expressions for $x,y$ as Weierstrass functions of the
lattice data,
\begin{align}
x  &  =4^{2/3}\wp(z,\tau)=4^{2/3}\left[  \frac{1}{z^{2}}+\sum_{(m,n)\neq
(0,0)}\left(  \frac{1}{(z+m+n\tau)^{2}}-\frac{1}{(m+n\tau)^{2}}\right)
\right]  ,\nonumber\\
y  &  =\wp(z,\tau)^{\prime}=\sum_{(m,n)\neq(0,0)}\frac{1}{(z+m+n\tau)^{3}},
\end{align}
one can see that the orientation-reversing involutions described above acts on
the coordinates of the Weierstrass model as
\begin{equation}
(x,y)\,\rightarrow\,(\bar{x},\bar{y})\quad\text{for}\,\rho_{\mathsf{F}_L}%
,\quad\text{and}\quad(x,y)\,\rightarrow\,(\bar{x},-\bar{y})\quad
\text{for}\,\rho_{\Omega}.
\end{equation}
Both of these are symmetries of the Weierstrass model \eqref{eq:weierstrassmodel} if the
constants $f,g$ are real.

\section{$\mathsf{Pin}^\pm$ Structures on the Klein Bottle}\label{app:kb}

Here we describe the eight possible choices of pin structures a spinor may have on a Klein bottle. Four of these are $\mathsf{Pin}^+$ where the reflection $\mathsf{R}$ acts such that $\mathsf{R}^2=1$, while the other four are $\mathsf{Pin}^-$ structures where $\mathsf{R}^2=(-1)^F$. We first start with a torus, $\hat{\Sigma}$, with real coordinates $(x,y)$ and complex structure $\tau=2iL$,\footnote{A similar construction equally works for $\tau=2iL+1/2$.} i.e., we have the identifications $x\sim x+1$ and $y\sim y+2L$. We obtain a Klein bottle $\Sigma$ after quotienting by
\begin{equation}\label{eq:kbquotient}
  \sigma: \; \; (x,y)\; \sim \; (-x,y+L).
\end{equation}
If $\hat{\Sigma}$ has Ramond spin structure along the $y$-direction, then the Klein bottle has $\mathsf{Pin}^+$ structure because the orientation reversal \ref{eq:kbquotient} acted on twice should act as $+1$ on spinors. Therefore, $\hat{\Sigma}$ with $(R,R)$ or $(NS,R)$ along the $(x,y)$ directions induce two separate $\mathsf{Pin}^+$ structures on $\Sigma$. The other two $\mathsf{Pin}^+$ structures are gotten by composing the quotient \ref{eq:kbquotient} by $(-1)^{F}$: this allows the Klein bottle to have $NS$ structure along its b-cycle. In total, the a-/b-cycle holonomies are $(\pm 1, \pm 1)$.

We now see that when $\hat{\Sigma}$ has Neveu-Schwarz spin structure along the $y$-direction, the Klein bottle has $\mathsf{Pin}^-$ structure. The $(R,NS)$ and $(NS,NS)$ spin structures of $\hat{\Sigma}$ induce two $\mathsf{Pin}^-$ structures. Since the holonomy of a spinor winding twice around the B-cycle of $\Sigma$ is $-1$, the holonomy of single winding is $\pm i$, the choice of sign of which depends on whether or not we compose $\sigma$ with $(-1)^F$. In summary, the a-/b-cycle holonomies are $(\pm 1, \pm i)$. For more details on pin structures of Klein bottles see \cite{Kaidi:2019tyf}.

Applying this knowledge to the F-theory geometry of $\mathsf{F}_L$ 7-branes, the a-cycle (the cycle whose direction flips) from above is one of circles in the elliptic fiber, $\mathbb{E}$, while the b-cycle is the angular direction of the $X_8$-$X_9$ plane. To have a valid M-theory realization, the space must have a $\mathsf{Pin}^+$ structure which from above we see requires spinors to be periodic along the spacial angular direction prior to the quotient. In other words, there must be a spin defect (or Ramond puncture) located at $Z=X_8+iX_9=0$ in order for the F-theory geometry to describe the $\mathsf{F}_L$ 7-brane.

\section{M-theory Lift of $ \mathsf{F}_{L,R}$}\label{app:mlift}
The purpose of this appendix is to show that the M-theory uplift of the IIA symmetries $\mathsf{F}_L$ and $\mathsf{F}_R$ are reflections along the M-theory circle, $S^1_{11}$, where the difference between the two arise from a choice of sign in the fermion action.

Looking first at the IIA massless bosonic fields, we have that
\begin{align}\label{eq:bostrans}
  & \textnormal{$+1$ under $(-1)^{F_L}$:} \; \; \; g_{\mu\nu}, B_2, \phi  \\
 & \textnormal{$-1$ under $(-1)^{F_L}$:} \; \; \; C_1, C_3.
\end{align}
This matches what we expect from M-theory after applying $dx_{11}\mapsto -dx_{11}$, and the fact that the M-theory 3-form transforms as a ``pseudo 3-form": $C^{M-th}_3\rightarrow -C^{M-th}_3$. As for the massless fermionic states, we need to understand how the $x_{11}$ reflection acts on the 11D gravitino, $\Psi_{MA}$. Our indices decompose as $M={\mu,11}$ and $A=(\alpha,\dot{\alpha})$, where $\alpha$ denotes $8_s$ in the $Spin(8)$ little group. Since 11D Majorana fermions must have $\mathsf{Pin}^+$-structure, the reflection must satisfy $\mathsf{R}^2_{11}=1$ so from the 11D kinetic term we have two choices
\begin{eqnarray}\label{eq:fermtrans}
  \mathsf{R}_{11}: & \Psi_{\mu A}\rightarrow \pm\Gamma_{11}\Psi_{\mu A} \\
              & \Psi_{11, A}\rightarrow \mp \Gamma_{11} \Psi_{11, A}.
\end{eqnarray}
For an 11D Majorana spinor, $Q_{A}$, we can choose a convention so that $\Gamma_{11}$ decomposes in the following way in 10D
\begin{equation}\label{eq:decomposing10D}
  \Gamma_{11}Q=\begin{pmatrix}
                 1 & 0 \\
                 0 & -1
               \end{pmatrix}\begin{pmatrix}
                              Q_{\alpha} \\
                              Q_{\dot{\alpha}}
                            \end{pmatrix},
\end{equation}
i.e., it becomes the 10D chirality matrix. Let's choose a $+$ sign in \ref{eq:fermtrans}, and name the reflection $\mathsf{R}^{+}_{11}$ accordingly, then the 10D gravitinos, $\psi_{\mu\alpha}:=\Psi_{\mu, \alpha}$, and dilatinos, $\lambda_{\alpha}:=\Psi_{11, \alpha}$, transform as:
\begin{eqnarray}
  \psi_{\mu \alpha}\mapsto +\psi_{\mu \alpha} & \psi_{\mu \dot{\alpha}}\mapsto -\psi_{\mu \dot{\alpha}} \\
  \lambda_\alpha\mapsto -\lambda_\alpha                        & \lambda_{\dot{\alpha}}\mapsto +\lambda_{\dot{\alpha}}.
\end{eqnarray}
This is exactly the expected behavior of $(-1)^{F_L}$ on the left moving R-NS states after noting that the dilatinos and gravitinos have opposite chirality in a given chiral/anti-chrial worldsheet sector as we can see from
\begin{equation}
  \mathbf{8}_v\times \mathbf{8}_c=\mathbf{8}_s+\mathbf{56}_c.
\end{equation}
This was nicely reflected in the fact that we had an additional minus sign from the dilatino coming from a gravitino with a leg along the reflected direction. Clearly if we chose the other sign \ref{eq:fermtrans}, we would have $(-1)^{F_R}$ so in summary
\begin{equation}\label{eq:refs}
  \mathsf{R}^{\pm}_{11}=(-1)^{F_{L,R}}.
\end{equation}

\section{Computing One-loop Chern-Simons Terms with Duality Bundles}\label{app:CS1loop}
In this Appendix, we provide the technical details behind the anomaly results in Section \ref{sec:ANOMO}. While the kind of calculations described here are not new (see  \cite{Bonetti:2011mw,Bonetti:2012fn,Bonetti:2013ela,Garcia-Etxebarria:2015ota,Corvilain:2017luj}), some of the detailed results are, and we also provide an introduction to computing them using modern techniques (using the formalism of $\eta$ invariants and the APS index theorem). Appendix \ref{sec:app1} describes the general computation to be carried out, and its relation to modern techniques. Appendix \ref{sec:app2} contains the application of the formalism to the R7-brane in the main text.

\subsection{One-loop Chern Simons Terms from the Anomaly Theory}\label{sec:app1}
Let us review the general setup we are interested in, using the oldest example of it, the Redlich anomaly \cite{Redlich:1983kn}. Consider a four-dimensional theory of a Weyl fermion, charged under a global $U(1)$ symmetry, compactified on an $S^1$ with periodic boundary conditions for fermions. Upon dimensional reduction, one obtains a 3D theory containing one massless Dirac fermion, and a KK tower of massive Dirac ones. The sign of the mass is correlated to the KK quantum number $k$. In the framework of effective field theory, one then integrates out the KK tower to produce a low-energy effective action producing a 3D theory including a massless Weyl fermion and the background current for the $U(1)$ symmetry.

As usual, integrating out the KK modes produces higher-derivative couplings in the effective action. But somewhat unusually, there is a triangle diagram contributing an (in general improperly quantized) Chern-Simons term to the the one-loop effective action, depending only on the sign of the mass \cite{Redlich:1983kn,Corvilain:2017luj}:
\begin{equation} S_{\text{3D}}\supset k_{\text{CS}}\int F\wedge A,\quad k_{\text{CS}}=\sum_{n\neq 0}  \,\text{sgn}\,  (n).\end{equation}
Here, $A$ is the connection for the background $U(1)$ symmetry, and $n$ is an index running over KK momentum. Regulating the sum via e.g. zeta-function regularization, one obtains
\begin{equation} k_{\text{CS}}\rightarrow\frac12,\end{equation}
which is the original result of Redlich \cite{Redlich:1983kn}. Since then, this formalism has been extended to the computation of CS terms in any number of dimensions, and also including the KK photon $A_{KK}$ in the reduction as well. For instance, as explained in \cite{Corvilain:2017luj}, including the KK photon in the example above also yields a one-loop term of the form
\begin{equation} S_{\text{3D}}\supset \tilde{k}_{\text{CS}}\int F_{KK}\wedge A,\quad \tilde{k}_{\text{CS}}=\sum_{n\neq 0} n \,\text{sgn}\,  (n)\,\rightarrow -\frac{1}{12}.\end{equation}
The general rule is that the contribution of the KK modes of a fermion of charge $q$ to a CS term of degree $m$ in $F_{KK}$ or $A_{KK}$ and $l$ in $F$ or $A$ will involve a regularized sum
\begin{equation} \sum_{n\neq0} q^l n^m \,\text{sgn}\,  (m_n),\end{equation}
where $m_n$ is the mass of the $n$-th KK mode. For instance, when antiperiodic boundary conditions are chosen, $m_n=n+\frac12$. Using these rules, and zeta function regularization, it is possible to compute the effective action coming from KK modes in all cases.

The CS terms described above have peculiarities, in particular, they are not properly quantized in general, unless anomalies in the parent theory cancel \cite{Corvilain:2017luj}. We would like to understand the features of the above calculation using the modern formalism of the anomaly theory for the fermions \cite{Witten:2015aba,Garcia-Etxebarria:2018ajm}. We will describe the setup more generally, so it can be of help to modern computations of one-loop effective actions in a wider class of problems. Consider  a $d+1$-dimensional theory, in which there is a chiral fermion which is potentially anomalous. As is described in many places \cite{Witten:2015aba,Yonekura:2016wuc,Garcia-Etxebarria:2018ajm}, the phase of the partition function is not well defined, and this variation is encoded in terms of a $(d+2)$ dimensional topological anomaly theory, which for a chiral fermion is an $\eta$-invariant \cite{Witten:2015aba}.   In general, the partition function of the fermion on a $d+1$ dimensional manifold $X_{d+1}$ is given by extending to an open $(d+2)$ manifold $\tilde{Y}_{d+2}$, which has $\partial\tilde{Y}_{d+2}=X_{d+1}$, and defining
 \begin{equation}Z_{X_{d+1}}= |Z_{X_{d+1}}| \, \exp(2\pi i\eta(\tilde{Y}_{d+2})) \,,\end{equation}
 where one uses generalized APS boundary conditions for the corresponding Dirac operator in $\tilde{Y}_{d+2}$. The anomaly is encoded in the fact that the partition function thus defined clearly depends on the choice of $\tilde{Y}_{d+2}$.  This dependence can be studied by considering the anomaly theory on a closed  $(d+2)$-dimensional manifold $Y_{d+2}$. In cases where this manifold is itself the boundary of a $(d+3)$-dimensional manifold $Z_{d+3}$, the value of the $\eta$ invariant can be computed by means of the APS index theorem, as
\begin{equation} \eta(Y_{d+2})=\text{Index}(Z_{d+3}) -\int_{Z_{d+3}} [\hat{A}(R)\text{ch}(F)]_{d+3} \,.\end{equation}

 In general, it is impossible to write down an expression for $\eta(\tilde{Y}_{d+2})$ as an integral of a local density, i.e. in terms of cohomology classes \cite{10.2307/1990912}. But we are interested only in the particular case where we are  doing KK reduction on a circle; this means that the manifold $X_{d+1}$ where the parent theory lives, as well as the closed manifolds $Y_{d+2}$ we use to study anomales, are both $S^1$ bundles
  \begin{equation} S^1\,\longrightarrow X_{d+1}\, \longrightarrow \, B_{d},\end{equation}
 \begin{equation} S^1\,\longrightarrow Y_{d+2}\, \longrightarrow \, B_{d+1},\end{equation}
 over some bases $B_d$, $B_{d+1}$ of $d$ and $(d+1)$ dimensions, respectively. In the limit where the fiber $S^1$ becomes very small (known in the math literature as the adiabatic limit \cite{10.2307/1990912,AIF_1994__44_1_249_0,10.2307/2155025}), it is actually possible to write down an expression for the $\eta$ invariant in terms of cohomology classes. The corresponding cohomological expression, including an ``eta form'' piece, was computed in \cite{AIF_1994__44_1_249_0} and reads
 \begin{equation} \eta(S^1\rightarrow B_{d+1})=\int_{B_d} [\hat{A}(R) \, \text{ch}(F) \hat{\eta}(F_{KK})]_d,\quad \hat{\eta}(F_{KK})=\frac{2\tanh(F_{KK}/2)-F_{KK}}{2F_{KK}\tanh(F_{KK}/2)}, \label{zhang}\end{equation}
 where the $\text{ch}(F)$ piece encodes the charges that the fermion may have under additional internal bundles, and $F_{KK}$ corresponds to the field strength of the KK photon. We remark that the calculation in \cite{AIF_1994__44_1_249_0} is only valid for the bounding Spin structure on the $S^1$ fiber\footnote{Although the bounding Spin structure depends on the Spin structure of the base and the precise choice of $S^1$ fiber bundle, it  does not necessarily coincide with the antiperiodic Spin structure on the fiber. This is the case for a product total space $S^1\times B$; but for instance, for the Hopf fibration $S^1\rightarrow S^3\rightarrow S^2$, the induced Spin structure is periodic \cite{Martelli:2012sz}.}. The expression for $\hat{\eta}$  above can be rewritten as
 \begin{equation} \hat{\eta}(F_{KK})=\sum_{n=0}^\infty \frac{\zeta(-n)}{n!} F_{KK}^{n},\end{equation}
  which makes contact with the expressions involving zeta-regularized sums described above; what Zhang computed in \cite{AIF_1994__44_1_249_0} can be interpreted as the generating function for the zeta-regularized sums that have appeared in the physics literature \cite{Bonetti:2011mw,Bonetti:2012fn,Bonetti:2013ela,Corvilain:2017luj}. In fact, the connection can be made even more precise: as explained in \cite{Witten:2015aba}, the partition function of a massive Dirac fermion in odd dimensions is itself naturally an $\eta$ invariant, which can be written in terms of an anomaly polynomial in one dimension more. Therefore, one naturally gets the expression
 \begin{equation} \eta(S^1\rightarrow B_{d+1})=\sum_{\text{KK modes}} \eta_{KK}= \sum_{\text{KK modes}} P_{d+1},\label{433}\end{equation}
 i.e. a sum over the $(d+1)$-dimensional part of anomaly polynomials of fermions, rather than the  $(d+3)$-dimensional part that was relevant in \cite{AIF_1994__44_1_249_0}. If the sums in \eq{433} are regularized, one ends up with the same result as in \eq{zhang}.

Let us illustrate this by going back to our original example, the Redlich anomaly, this time including gravitational terms. The contribution for a single 4D Weyl fermion reduced on a circle and charged under an additional $U(1)$ field (of field strength $F$) is
  \begin{equation} \eta(S^1\rightarrow B_{d+1})=\frac{1}{24}\int_{Y_4} (12 F^2-2F F_{KK} -p_1),\end{equation}
  which corresponds to an improperly quantized Chern-Simons term in 3D,
  \begin{equation}\frac{1}{24}\int_{Y_4} (12 F^2-2F F_{KK} -p_1)= \frac{1}{24}\int_{X_3} (12 F\wedge A -2A\wedge  F_{KK} -CS_3[g]),\end{equation}
  where $CS_3[g]$ is the gravitational Chern-Simons term. This reproduces the results of \cite{Corvilain:2017luj}.  The anomaly of the theory, encoded in the $Y_4$ dependence of the $\eta$ invariant due to the Dai-Freed theorem, gets mapped to the more ordinary fact that an improperly quantized Chern-Simons term depends on the extension to $Y_4$.  In a theory where local anomalies cancel, the theory is actually independent of the choice of $Y_4$, and this maps to properly quantized Chern-Simons terms. To see this, just use the APS index theorem for the total fermion content of the theory,
  \begin{equation} \eta(S^1\rightarrow B_{d+1})=\text{Index}-\int_{Z_{d+2}} \sum_{\text{fields}}[\hat{A}(R)\text{ch}(F)]_{d+3}=\text{Index} \,,\end{equation}
  where in the second equality we used the fact that local anomalies vanish. We see that the $\eta$ invariant is always integer quantized, which means, when written as a sum of Chern-Simons terms, that their coefficients must always be properly quantized.

We will also need the anomaly theory for Rarita-Schwinger and self-dual fields. The expression for the eta form of a self-dual field is\footnote{This can be obtained via the standard trick of regarding the self-dual field as a bispinor field, which is valid at the level of local anomalies; it can also be found in \cite{10.2307/2155025}.}
 \begin{equation} \hat{\eta}^{SD}= \frac{F_{KK}-\tanh(F_{KK})}{F_{KK}\tanh(F_{KK})},\end{equation}
 so that
  \begin{equation} \eta^{SD}(S^1\rightarrow B_d)=\frac12\int_{B_d} [L(R) \, \text{ch}(F) \hat{\eta}^{SD}(F_{KK})]_d. \end{equation}
Finally, for a Rarita-Schwinger field, we have
\begin{equation}\eta^{RS}= [\hat{A}(R)[\text{ch}(R)-d] \, \text{ch}(F) \hat{\eta}(F_{KK})]_d,\end{equation} which can be derived from the APS index theorem and the fact that the restriction of a $(d+2)$-dimensional RS fermion to the boundary produces both a RS and a Dirac fermion.  Using these  results, and as a warm-up, we can reproduce the main result in \cite{Bonetti:2013ela} for the 1-loop CS terms of F-theory on a circle, namely the anomaly theory
 \begin{equation} \frac{1}{24} \left(2 c_1 p_1 (T-12)+c_1^3 (T-9)\right).\end{equation}
Recovering these results is a nice consistency check of the expressions above. We can also recover this using \eq{433}; the anomaly polynomial for a Rarita-Schwinger and self-dual field KK mode are then
\begin{equation}P_{d+1}^{RS}= [\hat{A}(R)(\text{ch}(R)-d)\text{ch}(nF_{KK})]_{d+1},\quad P_{d+1}^{SD}=-\frac12 [L(R)\,\text{ch}(2nF_{KK})]_{d+1}  \end{equation}
respectively.
\subsection{Application to Circle Compactifications of IIB}\label{sec:app2}

We are now ready to move to our main arena of interest, compactifications of type IIB string theory on a circle. It is easiest to approach the problem using \eq{433} to write the Chern-Simons terms as a sum of 10D anomaly polynomials of the KK modes. The contribution of the $q$-th KK mode for the (complex) dilatini and gravitini, as well as self-dual fields, is
\begin{align}P_{\text{dilatini},q}=-\frac{F_{KK} q \left(-40 F_{KK}^2 p_1 q^2+48 F_{KK}^4 q^4+7 p_1^2-4 p_2\right)}{5760} \,\text{sgn}\,  (m^{\text{dilatino}}_q)\nonumber\\
P_{\text{Self-dual},q}=\frac{1}{90} F_{KK} q \left(-10 F_{KK}^2 p_1 q^2-6 F_{KK}^4 q^4+p_1^2-7 p_2\right) \,\text{sgn}\,  (m^{\text{SD}}_q)\nonumber\\
P_{\text{RS},q}=\frac{F_{KK} q \left(200 F_{KK}^2 p_1 q^2+144 F_{KK}^4 q^4+101 p_1^2-332 p_2\right)}{1920} \,\text{sgn}\,  (m^{\text{Gravitino}}_q).\end{align}
As in \eq{433}, the full anomaly theory is obtained by summing over KK modes. Since we are interested in compactifications of IIB string theory with a duality holonomy, the mass of each field will take the general expression
\begin{equation} m_q=q+\alpha,\end{equation}
for $\alpha$ dictated by the holonomy of each particular field under the duality transformation. In general, there is an ambiguity, since $\alpha$ is a real number, but the action of the duality bundle on fields only dictate the fractional part of $\alpha$. However, the expression \eq{433} for the $\eta$ invariant degenerates when $m_q$ crosses zero, as the $\eta$ invariant changes discontinuously (the $\eta$ invariant can jump as one crosses a fermion zero mode). So \eq{433} only really works if we restrict $\alpha$ to $\vert \alpha\vert<1$ for all fields in type IIB.  Regularizing the sums according to \cite{Corvilain:2017luj},
\begin{equation} \sum_q q\, \text{sgn}\,(q+\alpha)\,\rightarrow\, -\frac{1}{12}+\frac{\alpha^2}{2},\end{equation}
we obtain, for a circle with trivial holonomy (so that $\alpha^{\text{dilatino}}=\alpha^{\text{gravitino}}=\alpha^{\text{SD}}=0$)
\begin{equation} dCS_{IIB}=-\frac{1}{192} F_{KK} \left(p_1^2-4 p_2\right).\end{equation}
In particular, the terms proportional to $q^3$ and $q^5$ cancel out. Therefore, we recover a term $A_{KK} \wedge X_8$ in the 9d action, where the gravitational class $X_8$ is exactly right to match the dual IIA picture, as explained in \cite{Liu:2013Dna,Garcia-Etxebarria:2015ota}. Thus, our calculations are consistent with T-duality and reproduce the known one-loop topological terms of IIB in ten dimensions, and nothing more. Although this matching between IIA and IIB perspectives had appeared before in the literature, to our knowledge the explicit matching of $F_{KK}^3$, $F_{KK}^5$ terms seems to be new.

We can also look at other Mp$(2,\mathbb{Z})$ holonomies. For instance, for a holonomy of  $\hat{S}^2$ (corresponding to an $\mathfrak{so}(8)$ stack, so that $\alpha^{\text{dilatino}}=\alpha^{\text{gravitino}}=1/4, \, \alpha^{\text{SD}}=0$) we obtain a term
\begin{equation} dCS_{IIB}\supset-\frac{1}{256} F_{KK} \left(p_1^2-4 p_2\right).\label{e332}\end{equation}
Presumably, this is cancelled by anomaly inflow from the orientifold planes. This is a subtle topic, which is not fully understood; Reference \cite{Kim:2012wc} studied this in detail, and found some contributions that did not completely cancel. However, this reference did not take into account the one-loop contribution of loop modes as in \cite{Liu:2013Dna,Garcia-Etxebarria:2015ota}; it would be very interesting to see if the term \eq{e332} is enough to find a perfect match with the results of \cite{Kim:2012wc}, but this is beyond the scope of this paper. In this case, the terms proportional to $F_{KK}^3$ and $F_{KK}^5$ no longer cancel; but these contributions can be cancelled by inflow of RR fields, as in \cite{Kim:2012wc}. It would be interesting to study this in the future.

We are now ready for the calculation we set out to perform: the anomaly inflow on the R7-brane. For either R7-brane, we must now decompose dilatino and gravitino into two real fermions, each of which has $\alpha=0,1/2$, corresponding to the $\pm1$ eigenvalues of the action of the duality bundle in the AOB and DP backgrounds. And the self-dual field has $\alpha^{\text{SD}}=1/2$, corresponding to the fact that $\Omega$ and $(-1)^{F_L}$ flips its sign. Taking everything into account, we obtain an anomaly polynomial
\begin{equation}\frac{F_{KK} \left(10 F_{KK}^2 p_1+F_{KK}^4+7 p_1^2-4 p_2\right)}{11520}.\label{anomr7}\end{equation}
The last two terms coincide with the anomaly polynomial of a single Weyl fermion with charge $q=1/2$. We cannot single out this possibility, since, as we will discuss at the end of the Section, there are more possibilities that can cancel anomalies. However, if the fermion is really present, it has a natural physical interpretation; the DP and AOB backgrounds are supersymmetric, and the R7-brane is a boundary condition that spontaneously breaks that supersymmetry, so the fermion could be interpreted as a goldstino. It transforms in the same R-symmetry representation as the (spontaneously broken) supercharges. This calculation constitutes yet another line of reasoning indicating that these backgrounds break supersymmetry, complementing the arguments in Section \ref{sec:AFAR}.

There is one last caveat we must discuss. In \eq{anomr7}, the first two terms match the anomaly of a Weyl fermion, but this is not the case for the last two.
The difference, namely
\begin{equation} \frac{10 F_{KK}^3 p_1-F_{KK}^5}{5760}\label{anleft}\end{equation}
can be accounted for by additional terms not coming from the supergravity action can contribute to the anomaly inflow, as in \cite{Kim:2012wc}, and cancel this additional contribution. For instance, a term of the form
\begin{equation}\int B_2 X_6 + B_6 X_2\label{ewew4}\end{equation}
contributes to the anomaly a quantity proportional to \cite{Kim:2012wc}
\begin{equation} X_2 X_6 F_{KK},\end{equation}
and so it potentially has the right form to cancel \eq{anleft}. One might worry that couplings like
\begin{equation}\int C_0 X_8,\quad \text{or}\quad \int C_4X_4\label{e323}\end{equation}
could contribute, in principle, to pieces not of the form $X_2 X_8 F_{KK}$, potentially spoiling the anomaly match with Weyl fermions we found already for these terms. However, the R7 preserves one of the two symmetries $\Omega$ and $(-1)^{F_L}$. Either of these flips the sign of $C_0,C_4$, forbidding the terms above.

Finally, we also report on a partial consistency check of our picture, whose details we do not fully understand. Consider the case $\alpha^{\text{dilatino}}=\alpha^{\text{gravitino}}=1/2, \, \alpha^{\text{SD}}=0$, corresponding to a circle with \emph{antiperiodic} boundary conditions. A circle with antiperiodic boundary conditions is the boundary of a disk, so the corresponding domain wall is just a simple ``cigar'' geometry that can in principle be completely described in supergravity, unlike all other objects in this paper. A naive expectation would be that the Chern-Simons terms should vanish identically, since there is no singular brane. However, the result we find is instead exactly \emph{twice} of that in \eq{anomr7}, in spite of the fact that the combination of $\eta$ invariants that one must take is completely different. The fact that the two agree, which hinges on detailed cancelations, has a natural interpretation. As discussed in Section \ref{sec:AFAR}, $\Omega_1^{\text{Spin}-GL^+(2,\mathbb{Z})}=\mathbb{Z}_2 \times \mathbb{Z}_2$ is generated by either R7-brane, and so in particular there is a bordism taking two $\Omega$ or two $(-1)^{F_L}$ branes to the vacuum, i.e. a cigar geometry. As we have seen, configurations with R7-branes are particularly simple, and devoid of many of the RR couplings or axio-dilaton gradients that complicate inflow analysis for other setups; so the anomaly of the two branes together must simply add up to that of the cigar.

While our study of the anomaly of the cigar geometry has provided a nice consistency check, it raises questions of its own. The cigar is a completely smooth geometry in IIB, so what kind of fields can provide a boundary for the Chern-Simons terms? A piece of the answer is that, perhaps surprisingly, there are supergravity chiral zero modes (dilatino, gravitino\footnote{We might be alarmed that gravitini appear as localized fields in the cigar metric; but we must remember that this background does not solve the Euclidean equations of motion, it describes a time-dependent situation, and so, flat space representation theory is not relevant \cite{Montero:2020icj}. As an example consider compactifying IIB on a four-manifold which is the connected sum of $n>1$ copies of K3. We immediately get $n$ 6d gravitini, but this is fine, since the configuration is dynamical. }, and self-dual field) at the core of the cigar geometry. These zero modes are not directly detected in a simple application of the index theorem, but they may still contribute to the anomaly involving $U(1)_{KK}$. Some of these modes can be constructed explicitly using the conformal covariance of the Dirac, Rarita-Schwinger, and Laplace equations; but we have not been able to show that the combined contributions of the modes we have identified add up to twice \eq{anomr7}. It is possible we are missing additional zero modes that cannot be constructed easily with our techniques; in any case, we hope to return to this interesting question in the future.

\bibliographystyle{utphys}
\bibliography{r7brane}

\providecommand{\href}[2]{#2}\begingroup\raggedright\begin{thebibliography}{10}

\bibitem{Gaiotto:2014kfa}
D.~Gaiotto, A.~Kapustin, N.~Seiberg, and B.~Willett, ``{Generalized Global
  Symmetries},'' \href{http://dx.doi.org/10.1007/JHEP02(2015)172}{{\em JHEP}
  {\bfseries 02} (2015) 172}, \href{http://arxiv.org/abs/1412.5148}{{\ttfamily
  arXiv:1412.5148 [hep-th]}}.

\bibitem{McNamara:2019rup}
J.~McNamara and C.~Vafa, ``{Cobordism Classes and the Swampland},''
  \href{http://arxiv.org/abs/1909.10355}{{\ttfamily arXiv:1909.10355
  [hep-th]}}.

\bibitem{Montero:2020icj}
M.~Montero and C.~Vafa, ``{Cobordism Conjecture, Anomalies, and the String
  Lamppost Principle},'' \href{http://dx.doi.org/10.1007/JHEP01(2021)063}{{\em
  JHEP} {\bfseries 01} (2021) 063},
  \href{http://arxiv.org/abs/2008.11729}{{\ttfamily arXiv:2008.11729
  [hep-th]}}.

\bibitem{Dierigl:2020lai}
M.~Dierigl and J.~J. Heckman, ``{Swampland cobordism conjecture and non-Abelian
  duality groups},'' \href{http://dx.doi.org/10.1103/PhysRevD.103.066006}{{\em
  Phys. Rev. D} {\bfseries 103} no.~6, (2021) 066006},
  \href{http://arxiv.org/abs/2012.00013}{{\ttfamily arXiv:2012.00013
  [hep-th]}}.

\bibitem{Buratti:2021fiv}
G.~Buratti, J.~Calder\'on-Infante, M.~Delgado, and A.~M. Uranga, ``{Dynamical
  Cobordism and Swampland Distance Conjectures},''
  \href{http://dx.doi.org/10.1007/JHEP10(2021)037}{{\em JHEP} {\bfseries 10}
  (2021) 037}, \href{http://arxiv.org/abs/2107.09098}{{\ttfamily
  arXiv:2107.09098 [hep-th]}}.

\bibitem{Debray:2021vob}
A.~Debray, M.~Dierigl, J.~J. Heckman, and M.~Montero, ``{The anomaly that was
  not meant IIB},'' \href{http://arxiv.org/abs/2107.14227}{{\ttfamily
  arXiv:2107.14227 [hep-th]}}.

\bibitem{Blumenhagen:2021nmi}
R.~Blumenhagen and N.~Cribiori, ``{Open-closed correspondence of K-theory and
  cobordism},'' \href{http://dx.doi.org/10.1007/JHEP08(2022)037}{{\em JHEP}
  {\bfseries 08} (2022) 037}, \href{http://arxiv.org/abs/2112.07678}{{\ttfamily
  arXiv:2112.07678 [hep-th]}}.

\bibitem{Angius:2022aeq}
R.~Angius, J.~Calder\'on-Infante, M.~Delgado, J.~Huertas, and A.~M. Uranga,
  ``{At the end of the world: Local Dynamical Cobordism},''
  \href{http://dx.doi.org/10.1007/JHEP06(2022)142}{{\em JHEP} {\bfseries 06}
  (2022) 142}, \href{http://arxiv.org/abs/2203.11240}{{\ttfamily
  arXiv:2203.11240 [hep-th]}}.

\bibitem{Blumenhagen:2022mqw}
R.~Blumenhagen, N.~Cribiori, C.~Kneissl, and A.~Makridou, ``{Dynamical
  Cobordism of a Domain Wall and its Companion Defect 7-brane},''
  \href{http://arxiv.org/abs/2205.09782}{{\ttfamily arXiv:2205.09782
  [hep-th]}}.

\bibitem{Angius:2022mgh}
R.~Angius, M.~Delgado, and A.~M. Uranga, ``{Dynamical Cobordism and the
  Beginning of Time: Supercritical Strings and Tachyon Condensation},''
  \href{http://arxiv.org/abs/2207.13108}{{\ttfamily arXiv:2207.13108
  [hep-th]}}.

\bibitem{Blumenhagen:2022bvh}
R.~Blumenhagen, N.~Cribiori, C.~Kneissl, and A.~Makridou, ``{Dimensional
  Reduction of Cobordism and K-theory},''
  \href{http://arxiv.org/abs/2208.01656}{{\ttfamily arXiv:2208.01656
  [hep-th]}}.

\bibitem{Tachikawa:2018njr}
Y.~Tachikawa and K.~Yonekura, ``{Why are fractional charges of orientifolds
  compatible with Dirac quantization?},''
  \href{http://dx.doi.org/10.21468/SciPostPhys.7.5.058}{{\em SciPost Phys.}
  {\bfseries 7} no.~5, (2019) 058},
  \href{http://arxiv.org/abs/1805.02772}{{\ttfamily arXiv:1805.02772
  [hep-th]}}.

\bibitem{BIGKAHUNA}
A.~Debray, M.~Dierigl, J.~J. Heckman, and M.~Montero, ``{The Chronicles of
  IIBordia: Dualities, Bordisms, and the Swampland},'' {\em To Appear} .

\bibitem{Distler:2009ri}
J.~Distler, D.~S. Freed, and G.~W. Moore, ``{Orientifold Precis},''
  \href{http://arxiv.org/abs/0906.0795}{{\ttfamily arXiv:0906.0795 [hep-th]}}.

\bibitem{Dabholkar:1996pc}
A.~Dabholkar and J.~Park, ``{Strings on orientifolds},''
  \href{http://dx.doi.org/10.1016/0550-3213(96)00395-1}{{\em Nucl. Phys. B}
  {\bfseries 477} (1996) 701--714},
  \href{http://arxiv.org/abs/hep-th/9604178}{{\ttfamily arXiv:hep-th/9604178}}.

\bibitem{Hellerman:2005ja}
S.~Hellerman, ``{New type II string theories with sixteen supercharges},''
  \href{http://arxiv.org/abs/hep-th/0512045}{{\ttfamily arXiv:hep-th/0512045}}.

\bibitem{Aharony:2007du}
O.~Aharony, Z.~Komargodski, and A.~Patir, ``{The Moduli space and M(atrix)
  theory of 9d N=1 backgrounds of M/string theory},''
  \href{http://dx.doi.org/10.1088/1126-6708/2007/05/073}{{\em JHEP} {\bfseries
  05} (2007) 073}, \href{http://arxiv.org/abs/hep-th/0702195}{{\ttfamily
  arXiv:hep-th/0702195}}.

\bibitem{Schwarz:1982ec}
A.~S. Schwarz, ``{FIELD THEORIES WITH NO LOCAL CONSERVATION OF THE ELECTRIC
  CHARGE},'' \href{http://dx.doi.org/10.1016/0550-3213(82)90190-0}{{\em Nucl.
  Phys. B} {\bfseries 208} (1982) 141--158}.

\bibitem{Schwarz:1982zt}
A.~S. Schwarz and Y.~S. Tyupkin, ``{GRAND UNIFICATION AND MIRROR PARTICLES},''
  \href{http://dx.doi.org/10.1016/0550-3213(82)90265-6}{{\em Nucl. Phys. B}
  {\bfseries 209} (1982) 427--432}.

\bibitem{MonteroTalk}
M.~Montero, ``{Cobordisms and \reflectbox{F} theory},''.
  \url{{https://sites.google.com/view/strings-and-geometry-2022/program}}.

\bibitem{HeckmanTalk}
J.~J. Heckman, ``{Reflections on F-theory and the Swampland Cobordism
  Conjecture},''.
  \url{{http://www.maths.liv.ac.uk/stringpheno2022/program.html}}.

\bibitem{Witten:2016cio}
E.~Witten, ``{The ``Parity'' Anomaly On An Unorientable Manifold},''
  \href{http://dx.doi.org/10.1103/PhysRevB.94.195150}{{\em Phys. Rev. B}
  {\bfseries 94} no.~19, (2016) 195150},
  \href{http://arxiv.org/abs/1605.02391}{{\ttfamily arXiv:1605.02391
  [hep-th]}}.

\bibitem{Vafa:1996xn}
C.~Vafa, ``{Evidence for F theory},''
  \href{http://dx.doi.org/10.1016/0550-3213(96)00172-1}{{\em Nucl. Phys. B}
  {\bfseries 469} (1996) 403--418},
  \href{http://arxiv.org/abs/hep-th/9602022}{{\ttfamily arXiv:hep-th/9602022}}.

\bibitem{Morrison:1996na}
D.~R. Morrison and C.~Vafa, ``{Compactifications of F-Theory on Calabi-Yau
  Threefolds -- I},''
  \href{http://dx.doi.org/10.1016/0550-3213(96)00242-8}{{\em Nucl. Phys. B}
  {\bfseries 473} (1996) 74--92},
  \href{http://arxiv.org/abs/hep-th/9602114}{{\ttfamily arXiv:hep-th/9602114}}.

\bibitem{Morrison:1996pp}
D.~R. Morrison and C.~Vafa, ``{Compactifications of F-Theory on Calabi-Yau
  Threefolds -- II},''
  \href{http://dx.doi.org/10.1016/0550-3213(96)00369-0}{{\em Nucl. Phys. B}
  {\bfseries 476} (1996) 437--469},
  \href{http://arxiv.org/abs/hep-th/9603161}{{\ttfamily arXiv:hep-th/9603161}}.

\bibitem{Pantev:2016nze}
T.~Pantev and E.~Sharpe, ``{Duality group actions on fermions},''
  \href{http://dx.doi.org/10.1007/JHEP11(2016)171}{{\em JHEP} {\bfseries 11}
  (2016) 171}, \href{http://arxiv.org/abs/1609.00011}{{\ttfamily
  arXiv:1609.00011 [hep-th]}}.

\bibitem{Morrison:2012np}
D.~R. Morrison and W.~Taylor, ``{Classifying bases for 6D F-theory models},''
  \href{http://dx.doi.org/10.2478/s11534-012-0065-4}{{\em Central Eur. J.
  Phys.} {\bfseries 10} (2012) 1072--1088},
  \href{http://arxiv.org/abs/1201.1943}{{\ttfamily arXiv:1201.1943 [hep-th]}}.

\bibitem{Garcia-Etxebarria:2015wns}
I.~Garcia-Etxebarria and D.~Regalado, ``{$ \mathcal{N}=3 $ four dimensional
  field theories},'' \href{http://dx.doi.org/10.1007/JHEP03(2016)083}{{\em
  JHEP} {\bfseries 03} (2016) 083},
  \href{http://arxiv.org/abs/1512.06434}{{\ttfamily arXiv:1512.06434
  [hep-th]}}.

\bibitem{Aharony:2016kai}
O.~Aharony and Y.~Tachikawa, ``{S-folds and 4d $\mathcal{N}=3$ superconformal
  field theories},'' \href{http://dx.doi.org/10.1007/JHEP06(2016)044}{{\em
  JHEP} {\bfseries 06} (2016) 044},
  \href{http://arxiv.org/abs/1602.08638}{{\ttfamily arXiv:1602.08638
  [hep-th]}}.

\bibitem{McNamara:2021cuo}
J.~McNamara, ``{Gravitational Solitons and Completeness},''
  \href{http://arxiv.org/abs/2108.02228}{{\ttfamily arXiv:2108.02228
  [hep-th]}}.

\bibitem{Dasgupta:1996ij}
K.~Dasgupta and S.~Mukhi, ``{F-Theory at Constant Coupling},''
  \href{http://dx.doi.org/10.1016/0370-2693(96)00875-1}{{\em Phys. Lett. B}
  {\bfseries 385} (1996) 125--131},
  \href{http://arxiv.org/abs/hep-th/9606044}{{\ttfamily arXiv:hep-th/9606044}}.

\bibitem{Heckman:2017uxe}
J.~J. Heckman and L.~Tizzano, ``{6D Fractional Quantum Hall Effect},''
  \href{http://dx.doi.org/10.1007/JHEP05(2018)120}{{\em JHEP} {\bfseries 05}
  (2018) 120}, \href{http://arxiv.org/abs/1708.02250}{{\ttfamily
  arXiv:1708.02250 [hep-th]}}.

\bibitem{Heckman:2018mxl}
J.~J. Heckman, C.~Lawrie, L.~Lin, and G.~Zoccarato, ``{F-theory and Dark
  Energy},'' \href{http://dx.doi.org/10.1002/prop.201900057}{{\em Fortsch.
  Phys.} {\bfseries 67} no.~10, (2019) 1900057},
  \href{http://arxiv.org/abs/1811.01959}{{\ttfamily arXiv:1811.01959
  [hep-th]}}.

\bibitem{Heckman:2022peq}
J.~J. Heckman, A.~Joyce, J.~Sakstein, and M.~Trodden, ``{Exploring
  $\boldsymbol{2+2}$ Answers to $\boldsymbol{3+1}$ Questions},''
  \href{http://arxiv.org/abs/2208.02267}{{\ttfamily arXiv:2208.02267
  [hep-th]}}.

\bibitem{Dabholkar:1997zd}
A.~Dabholkar, ``{Lectures on orientifolds and duality},'' in {\em {ICTP Summer
  School in High-Energy Physics and Cosmology}}, pp.~128--191.
\newblock 6, 1997.
\newblock \href{http://arxiv.org/abs/hep-th/9804208}{{\ttfamily
  arXiv:hep-th/9804208}}.

\bibitem{Frau:1999qs}
M.~Frau, L.~Gallot, A.~Lerda, and P.~Strigazzi, ``{Stable non-BPS D-branes in
  Type I String Theory},''
  \href{http://dx.doi.org/10.1016/S0550-3213(99)00624-0}{{\em Nucl. Phys. B}
  {\bfseries 564} (2000) 60--85},
  \href{http://arxiv.org/abs/hep-th/9903123}{{\ttfamily arXiv:hep-th/9903123}}.

\bibitem{Sen:1999mg}
A.~Sen, ``{Non-BPS States and Branes in String Theory},'' in {\em {Advanced
  School on Supersymmetry in the Theories of Fields, Strings and Branes}},
  pp.~187--234.
\newblock 1, 1999.
\newblock \href{http://arxiv.org/abs/hep-th/9904207}{{\ttfamily
  arXiv:hep-th/9904207}}.

\bibitem{Adams:2001sv}
A.~Adams, J.~Polchinski, and E.~Silverstein, ``{Don't panic! Closed string
  tachyons in ALE space-times},''
  \href{http://dx.doi.org/10.1088/1126-6708/2001/10/029}{{\em JHEP} {\bfseries
  10} (2001) 029}, \href{http://arxiv.org/abs/hep-th/0108075}{{\ttfamily
  arXiv:hep-th/0108075}}.

\bibitem{Witten:1998cd}
E.~Witten, ``{D-branes and K-theory},''
  \href{http://dx.doi.org/10.1088/1126-6708/1998/12/019}{{\em JHEP} {\bfseries
  12} (1998) 019}, \href{http://arxiv.org/abs/hep-th/9810188}{{\ttfamily
  arXiv:hep-th/9810188}}.

\bibitem{Loaiza-Brito:2001yer}
O.~Loaiza-Brito and A.~M. Uranga, ``{The Fate of the type I nonBPS D7-brane},''
  \href{http://dx.doi.org/10.1016/S0550-3213(01)00505-3}{{\em Nucl. Phys. B}
  {\bfseries 619} (2001) 211--231},
  \href{http://arxiv.org/abs/hep-th/0104173}{{\ttfamily arXiv:hep-th/0104173}}.

\bibitem{Weigand:2018rez}
T.~Weigand, ``{F-theory},'' {\em PoS} {\bfseries TASI2017} (2018) 016,
  \href{http://arxiv.org/abs/1806.01854}{{\ttfamily arXiv:1806.01854
  [hep-th]}}.

\bibitem{Cvetic:2022uuu}
M.~Cveti\v{c}, M.~Dierigl, L.~Lin, and H.~Y. Zhang, ``{All eight- and
  nine-dimensional string vacua from junctions},''
  \href{http://dx.doi.org/10.1103/PhysRevD.106.026007}{{\em Phys. Rev. D}
  {\bfseries 106} no.~2, (2022) 026007},
  \href{http://arxiv.org/abs/2203.03644}{{\ttfamily arXiv:2203.03644
  [hep-th]}}.

\bibitem{Cvetic:2021sxm}
M.~Cvetic, M.~Dierigl, L.~Lin, and H.~Y. Zhang, ``{Higher-form symmetries and
  their anomalies in M-/F-theory duality},''
  \href{http://dx.doi.org/10.1103/PhysRevD.104.126019}{{\em Phys. Rev. D}
  {\bfseries 104} no.~12, (2021) 126019},
  \href{http://arxiv.org/abs/2106.07654}{{\ttfamily arXiv:2106.07654
  [hep-th]}}.

\bibitem{Callan:1984sa}
C.~G. Callan, Jr. and J.~A. Harvey, ``{Anomalies and Fermion Zero Modes on
  Strings and Domain Walls},''
  \href{http://dx.doi.org/10.1016/0550-3213(85)90489-4}{{\em Nucl. Phys. B}
  {\bfseries 250} (1985) 427--436}.

\bibitem{Montero:2022vva}
M.~Montero and H.~Parra~de Freitas, ``{New Supersymmetric String Theories from
  Discrete Theta Angles},'' \href{http://arxiv.org/abs/2209.03361}{{\ttfamily
  arXiv:2209.03361 [hep-th]}}.

\bibitem{Sethi:2013hra}
S.~Sethi, ``{A New String in Ten Dimensions?},''
  \href{http://dx.doi.org/10.1007/JHEP09(2013)149}{{\em JHEP} {\bfseries 09}
  (2013) 149}, \href{http://arxiv.org/abs/1304.1551}{{\ttfamily arXiv:1304.1551
  [hep-th]}}.

\bibitem{Witten:1981gj}
E.~Witten, ``{Instability of the Kaluza-Klein Vacuum},''
  \href{http://dx.doi.org/10.1016/0550-3213(82)90007-4}{{\em Nucl. Phys. B}
  {\bfseries 195} (1982) 481--492}.

\bibitem{Horava:1996ma}
P.~Horava and E.~Witten, ``{Eleven-dimensional supergravity on a manifold with
  boundary},'' \href{http://dx.doi.org/10.1016/0550-3213(96)00308-2}{{\em Nucl.
  Phys. B} {\bfseries 475} (1996) 94--114},
  \href{http://arxiv.org/abs/hep-th/9603142}{{\ttfamily arXiv:hep-th/9603142}}.

\bibitem{Witten:1982fp}
E.~Witten, ``{An SU(2) Anomaly},''
  \href{http://dx.doi.org/10.1016/0370-2693(82)90728-6}{{\em Phys. Lett. B}
  {\bfseries 117} (1982) 324--328}.

\bibitem{Cordova:2019bsd}
C.~C\'ordova and K.~Ohmori, ``{Anomaly Obstructions to Symmetry Preserving
  Gapped Phases},'' \href{http://arxiv.org/abs/1910.04962}{{\ttfamily
  arXiv:1910.04962 [hep-th]}}.

\bibitem{Witten:2015aba}
E.~Witten, ``{Fermion Path Integrals And Topological Phases},''
  \href{http://dx.doi.org/10.1103/RevModPhys.88.035001}{{\em Rev. Mod. Phys.}
  {\bfseries 88} no.~3, (2016) 035001},
  \href{http://arxiv.org/abs/1508.04715}{{\ttfamily arXiv:1508.04715
  [cond-mat.mes-hall]}}.

\bibitem{ortin2004gravity}
T.~Ort{\'\i}n, {\em Gravity and Strings}.
\newblock Cambridge Monographs on Mathematical Physics. Cambridge University
  Press, 2004.

\bibitem{Lee:2022spd}
Y.~Lee and K.~Yonekura, ``{Global anomalies in 8d supergravity},''
  \href{http://dx.doi.org/10.1007/JHEP07(2022)125}{{\em JHEP} {\bfseries 07}
  (2022) 125}, \href{http://arxiv.org/abs/2203.12631}{{\ttfamily
  arXiv:2203.12631 [hep-th]}}.

\bibitem{Wang:2022ucy}
J.~Wang and Y.-Z. You, ``{Symmetric Mass Generation},''
  \href{http://dx.doi.org/10.3390/sym14071475}{{\em Symmetry} {\bfseries 14}
  no.~7, (2022) 1475}, \href{http://arxiv.org/abs/2204.14271}{{\ttfamily
  arXiv:2204.14271 [cond-mat.str-el]}}.

\bibitem{Martucci:2022krl}
L.~Martucci, N.~Risso, and T.~Weigand, ``{Quantum Gravity Bounds on N=1
  Effective Theories in Four Dimensions},''
  \href{http://arxiv.org/abs/2210.10797}{{\ttfamily arXiv:2210.10797
  [hep-th]}}.

\bibitem{Johnson:2003glb}
C.~V. Johnson, \href{http://dx.doi.org/10.1017/CBO9780511606540}{{\em
  {D-branes}}}.
\newblock Cambridge Monographs on Mathematical Physics. Cambridge University
  Press, 2005.

\bibitem{Witten:1996md}
E.~Witten, ``{On flux quantization in M theory and the effective action},''
  \href{http://dx.doi.org/10.1016/S0393-0440(96)00042-3}{{\em J. Geom. Phys.}
  {\bfseries 22} (1997) 1--13},
  \href{http://arxiv.org/abs/hep-th/9609122}{{\ttfamily arXiv:hep-th/9609122}}.

\bibitem{Bonetti:2013fma}
F.~Bonetti, T.~W. Grimm, and T.~G. Pugh, ``{Non-Supersymmetric F-Theory
  Compactifications on $Spin(7)$ Manifolds},''
  \href{http://dx.doi.org/10.1007/JHEP01(2014)112}{{\em JHEP} {\bfseries 01}
  (2014) 112}, \href{http://arxiv.org/abs/1307.5858}{{\ttfamily arXiv:1307.5858
  [hep-th]}}.

\bibitem{Bonetti:2013nka}
F.~Bonetti, T.~W. Grimm, E.~Palti, and T.~G. Pugh, ``{F-Theory on $Spin(7)$
  Manifolds: Weak-Coupling Limit},''
  \href{http://dx.doi.org/10.1007/JHEP02(2014)076}{{\em JHEP} {\bfseries 02}
  (2014) 076}, \href{http://arxiv.org/abs/1309.2287}{{\ttfamily arXiv:1309.2287
  [hep-th]}}.

\bibitem{Heckman:2019dsj}
J.~J. Heckman, C.~Lawrie, L.~Lin, J.~Sakstein, and G.~Zoccarato, ``{Pixelated
  Dark Energy},'' \href{http://dx.doi.org/10.1002/prop.201900071}{{\em Fortsch.
  Phys.} {\bfseries 67} no.~11, (2019) 1900071},
  \href{http://arxiv.org/abs/1901.10489}{{\ttfamily arXiv:1901.10489
  [hep-th]}}.

\bibitem{Cvetic:2020piw}
M.~Cveti\v{c}, J.~J. Heckman, T.~B. Rochais, E.~Torres, and G.~Zoccarato,
  ``{Geometric Unification of Higgs Bundle Vacua},''
  \href{http://dx.doi.org/10.1103/PhysRevD.102.106012}{{\em Phys. Rev. D}
  {\bfseries 102} no.~10, (2020) 106012},
  \href{http://arxiv.org/abs/2003.13682}{{\ttfamily arXiv:2003.13682
  [hep-th]}}.

\bibitem{Cvetic:2021maf}
M.~Cveti\v{c}, J.~J. Heckman, E.~Torres, and G.~Zoccarato, ``{Reflections on
  the matter of 3D N=1 vacua and local Spin(7) compactifications},''
  \href{http://dx.doi.org/10.1103/PhysRevD.105.026008}{{\em Phys. Rev. D}
  {\bfseries 105} no.~2, (2022) 026008},
  \href{http://arxiv.org/abs/2107.00025}{{\ttfamily arXiv:2107.00025
  [hep-th]}}.

\bibitem{Nahm:1977tg}
W.~Nahm, ``{Supersymmetries and their Representations},''
  \href{http://dx.doi.org/10.1016/0550-3213(78)90218-3}{{\em Nucl. Phys. B}
  {\bfseries 135} (1978) 149}.

\bibitem{Gubser:1996de}
S.~S. Gubser, I.~R. Klebanov, and A.~W. Peet, ``{Entropy and temperature of
  black 3-branes},'' \href{http://dx.doi.org/10.1103/PhysRevD.54.3915}{{\em
  Phys. Rev. D} {\bfseries 54} (1996) 3915--3919},
  \href{http://arxiv.org/abs/hep-th/9602135}{{\ttfamily arXiv:hep-th/9602135}}.

\bibitem{Gubser:1997yh}
S.~S. Gubser, I.~R. Klebanov, and A.~A. Tseytlin, ``{String theory and
  classical absorption by three-branes},''
  \href{http://dx.doi.org/10.1016/S0550-3213(97)00325-8}{{\em Nucl. Phys. B}
  {\bfseries 499} (1997) 217--240},
  \href{http://arxiv.org/abs/hep-th/9703040}{{\ttfamily arXiv:hep-th/9703040}}.

\bibitem{Arbys}
 \url{{https://www.arbys.com/menu/roast-beef/}}.

\bibitem{Kaidi:2019tyf}
J.~Kaidi, J.~Parra-Martinez, Y.~Tachikawa, and w.~a. m. a. b.~A. Debray,
  ``{Topological Superconductors on Superstring Worldsheets},''
  \href{http://dx.doi.org/10.21468/SciPostPhys.9.1.010}{{\em SciPost Phys.}
  {\bfseries 9} (2020) 10}, \href{http://arxiv.org/abs/1911.11780}{{\ttfamily
  arXiv:1911.11780 [hep-th]}}.

\bibitem{Bonetti:2011mw}
F.~Bonetti and T.~W. Grimm, ``{Six-dimensional (1,0) effective action of
  F-theory via M-theory on Calabi-Yau threefolds},''
  \href{http://dx.doi.org/10.1007/JHEP05(2012)019}{{\em JHEP} {\bfseries 05}
  (2012) 019}, \href{http://arxiv.org/abs/1112.1082}{{\ttfamily arXiv:1112.1082
  [hep-th]}}.

\bibitem{Bonetti:2012fn}
F.~Bonetti, T.~W. Grimm, and S.~Hohenegger, ``{A Kaluza-Klein inspired action
  for chiral p-forms and their anomalies},''
  \href{http://dx.doi.org/10.1016/j.physletb.2013.02.041}{{\em Phys. Lett. B}
  {\bfseries 720} (2013) 424--427},
  \href{http://arxiv.org/abs/1206.1600}{{\ttfamily arXiv:1206.1600 [hep-th]}}.

\bibitem{Bonetti:2013ela}
F.~Bonetti, T.~W. Grimm, and S.~Hohenegger, ``{One-loop Chern-Simons terms in
  five dimensions},'' \href{http://dx.doi.org/10.1007/JHEP07(2013)043}{{\em
  JHEP} {\bfseries 07} (2013) 043},
  \href{http://arxiv.org/abs/1302.2918}{{\ttfamily arXiv:1302.2918 [hep-th]}}.

\bibitem{Garcia-Etxebarria:2015ota}
I.~Garcia-Etxebarria, M.~Montero, and A.~M. Uranga, ``{Closed tachyon solitons
  in type II string theory},''
  \href{http://dx.doi.org/10.1002/prop.201500029}{{\em Fortsch. Phys.}
  {\bfseries 63} (2015) 571--595},
  \href{http://arxiv.org/abs/1505.05510}{{\ttfamily arXiv:1505.05510
  [hep-th]}}.

\bibitem{Corvilain:2017luj}
P.~Corvilain, T.~W. Grimm, and D.~Regalado, ``{Chiral anomalies on a circle and
  their cancellation in F-theory},''
  \href{http://dx.doi.org/10.1007/JHEP04(2018)020}{{\em JHEP} {\bfseries 04}
  (2018) 020}, \href{http://arxiv.org/abs/1710.07626}{{\ttfamily
  arXiv:1710.07626 [hep-th]}}.

\bibitem{Redlich:1983kn}
A.~N. Redlich, ``{Gauge Noninvariance and Parity Violation of Three-Dimensional
  Fermions},'' \href{http://dx.doi.org/10.1103/PhysRevLett.52.18}{{\em Phys.
  Rev. Lett.} {\bfseries 52} (1984) 18}.

\bibitem{Garcia-Etxebarria:2018ajm}
I.~Garcia-Etxebarria and M.~Montero, ``{Dai-Freed anomalies in particle
  physics},'' \href{http://dx.doi.org/10.1007/JHEP08(2019)003}{{\em JHEP}
  {\bfseries 08} (2019) 003}, \href{http://arxiv.org/abs/1808.00009}{{\ttfamily
  arXiv:1808.00009 [hep-th]}}.

\bibitem{Yonekura:2016wuc}
K.~Yonekura, ``{Dai-Freed theorem and topological phases of matter},''
  \href{http://dx.doi.org/10.1007/JHEP09(2016)022}{{\em JHEP} {\bfseries 09}
  (2016) 022}, \href{http://arxiv.org/abs/1607.01873}{{\ttfamily
  arXiv:1607.01873 [hep-th]}}.

\bibitem{10.2307/1990912}
J.-M. Bismut and J.~Cheeger, ``Eta-invariants and their adiabatic limits,''
  {\em Journal of the American Mathematical Society} {\bfseries 2} no.~1,
  (1989) 33--70.

\bibitem{AIF_1994__44_1_249_0}
W.~Zhang, ``Circle bundles, adiabatic limits of $\eta $-invariants and
  {Rokhlin} congruences,'' \href{http://dx.doi.org/10.5802/aif.1396}{{\em
  Annales de l'Institut Fourier} {\bfseries 44} no.~1, (1994) 249--270}.

\bibitem{10.2307/2155025}
X.~Dai and W.~Zhang, ``Circle bundles and the kreck-stolz invariant,'' {\em
  Transactions of the American Mathematical Society} {\bfseries 347} no.~9,
  (1995) 3587--3593.

\bibitem{Martelli:2012sz}
D.~Martelli, A.~Passias, and J.~Sparks, ``{The supersymmetric NUTs and bolts of
  holography},'' \href{http://dx.doi.org/10.1016/j.nuclphysb.2013.04.026}{{\em
  Nucl. Phys. B} {\bfseries 876} (2013) 810--870},
  \href{http://arxiv.org/abs/1212.4618}{{\ttfamily arXiv:1212.4618 [hep-th]}}.

\bibitem{Liu:2013Dna}
J.~T. Liu and R.~Minasian, ``{Higher-derivative couplings in string theory:
  dualities and the B-field},''
  \href{http://dx.doi.org/10.1016/j.nuclphysb.2013.06.002}{{\em Nucl. Phys. B}
  {\bfseries 874} (2013) 413--470},
  \href{http://arxiv.org/abs/1304.3137}{{\ttfamily arXiv:1304.3137 [hep-th]}}.

\bibitem{Kim:2012wc}
H.~Kim and P.~Yi, ``{D-brane anomaly inflow revisited},''
  \href{http://dx.doi.org/10.1007/JHEP02(2012)012}{{\em JHEP} {\bfseries 02}
  (2012) 012}, \href{http://arxiv.org/abs/1201.0762}{{\ttfamily arXiv:1201.0762
  [hep-th]}}.

\end{thebibliography}\endgroup

\end{document}